\def\simlt{\lower.5ex\hbox{$\; \buildrel < \over \sim \;$}}
\def\simgt{\lower.5ex\hbox{$\; \buildrel > \over \sim \;$}}
\RequirePackage{lineno} 
 \documentclass[preprint2]{emulateapj}
 \usepackage{graphicx}
 \usepackage[export]{adjustbox}
 \usepackage{float}

\newcommand{\myemail}{mrline@asu.edu}
\slugcomment{Submitted to the Astrophysical Journal }
\shorttitle{Brown Dwarf Retrieval}
\shortauthors{Line et al.}

\newcommand{\Spitzer}{{\sl Spitzer}}

\newcommand{\HST}{{\sl HST}}

\newcommand{\Mjup}{\mbox{$M_{\rm Jup}$}}

\newcommand{\etal}{et al.}

\newcommand{\Teff}{\mbox{$T_{\rm eff}$}}
\newcommand{\logg}{\mbox{$\log(g)$}}

\newcommand{\Lp}{\mbox{${L^\prime}$}}

\def\lesssim{\mathrel{\hbox{\rlap{\hbox{%
 \lower4pt\hbox{$\sim$}}}\hbox{$<$}}}}
\def\gtrsim{\mathrel{\hbox{\rlap{\hbox{%
 \lower4pt\hbox{$\sim$}}}\hbox{$>$}}}}
\def\wig#1{\mathrel{\hbox{\hbox to 0pt{   
  \lower.5ex\hbox{$\sim$}\hss}\raise.4ex\hbox{$#1$}}}}

\begin{document}
\title{Uniform Atmospheric Retrieval Analysis of Ultracool Dwarfs \\II:  Properties of 11 T-dwarfs}
\author{Michael R. Line}
\affil{School of Earth \& Space Exploration, Arizona State University,  Tempe AZ 85287, USA}
\author{Mark S. Marley}
\affil{NASA Ames Research Center, Mail Stop 245-3; Moffett Field, CA 94035, USA}
\author{Michael C. Liu}
\affil{Institute for Astronomy, University of Hawaii, 2680 Woodlawn Drive, Honolulu, HI 96822, USA}
\author{Ben Burningham}
\affil{ Centre for Astrophysics Research, School of Physics, Astronomy and Mathematics, University of Hertfordshire, Hatfield AL10 9AB, UK}
\affil{Marie Curie Fellow}
\author{Caroline V.  Morley}
\affil{ Department of Astronomy, Harvard University, Cambridge, MA 02138, USA}
\affil{NASA Sagan Fellow}
\author{Natalie R. Hinkel}
\affil{Department of Physics \& Astronomy, Vanderbilt University, Nashville, TN 37235, USA}
\author{Johanna Teske}
\affil{Observatories of the Carnegie Institution for Science, 813 Santa Barbara St., Pasadena, CA 91101}
\affil{Carnegie Origins Fellow, jointly appointed by Carnegie DTM \& Carnegie Observatories}
\author{Jonathan J. Fortney}
\affil{Department of Astronomy and Astrophysics, University of California, Santa Cruz, CA 95064}
\author{Richard Freedman}
\affil{SETI Institute,Mountain View, CA, USA }
\affil{NASA Ames Research Center, Mail Stop 245-3; Moffett Field, CA 94035, USA}
\author{Roxana Lupu}
\affil{BAER Institute/NASA Ames Research Center, Mail Stop 245-3; Moffett Field, CA 94035, USA}
\altaffiltext{1}{Correspondence to be directed to \myemail}

\begin{abstract}
Brown dwarf spectra are rich in information revealing of the chemical and physical processes operating in their atmospheres. We apply a recently developed atmospheric retrieval tool to an ensemble of late T-dwarf  (600-800K) near-infrared (1-2.5 $\mu$m) spectra.  With these spectra we are  able to directly constrain the molecular abundances {   for the first time} of H$_2$O, CH$_4$, CO, CO$_2$, NH$_3$, H$_2$S, and Na+K, surface gravity, effective temperature, thermal structure,  photometric radius, and cloud optical depths. We find that ammonia, water, methane, and the alkali metals are present and that {   their abundances are} well constrained in all 11 objects.  We find no significant trend in the water, methane, or ammonia abundances with temperature, but find a very strong ($>$25$\sigma$) decreasing trend in the alkali metal abundances with decreasing effective temperature, indicative of alkali rainout.  {   As expected from previous work}, we also find little evidence for optically thick clouds.  With the methane and water abundances, we derive the intrinsic atmospheric metallicity and carbon-to-oxygen ratios. We find in our sample that metallicities are typically sub-solar (-0.4 $<$ [M/H] $<$ 0.1 dex) and carbon-to-oxygen ratios are somewhat super-solar (0.4 $<$ C/O$ <$ 1.2), different than expectations from the local stellar population. We also find that the retrieved vertical thermal profiles are consistent with radiative equilibrium over the photospheric regions.  Finally, we find that our retrieved effective temperatures are lower than previous inferences for some objects and that some of our radii are larger than expectations from evolutionary models, possibly indicative of un-resolved binaries.  This investigation and methodology represents a {   new and powerful} paradigm for using spectra to the determine the fundamental chemical and physical processes governing cool brown dwarf atmospheres.
\end{abstract}

\section{Introduction}\label{sec:intro}
Brown dwarf spectra contain a wealth of information about the physical and chemical processes occurring in their atmospheres and possible formation avenues. Such processes include, but are not limited to; vertical atmospheric energy balance, disequilibrium chemistry, cloud formation/distribution and subsequent impact on molecular composition, variability,  and contain both chemical and physical links to their formation and evolution (see Marley \& Robinson 2015 review).  The main diagnostic quantities that can be obtained directly from a spectrum are the vertical thermal structure, molecular abundances, spectroscopic radius, and cloud properties. Deriving these diagnostic quantities is no simple task.   Progress in characterization always involves the interplay between observational and modeling efforts. It is from the comparison with models that we aim to understand the underlying physics and chemistry in substellar atmospheres.

The classic modeling approach for interpreting brown dwarf spectra is to generate a set of predictive self-consistent radiative- convective-thermochemical equilibrium models which can be compared to the data (Allard et al. 1996, Marley et al. 1996, Tsuji et al. 1996, Burrows et al. 2001).  Grids of model spectra are typically generated by varying only a few parameters, usually gravity  and effective temperature with all of the physics and chemistry in the atmospheres assumed (e.g., self-consistent radiative equilibrium plus equilibrium chemistry is used to predict the thermal structures and molecular abundances). These models are then scanned or interpolated along these parameters via a chi-squared search until a best fit, or range of best fit parameters are found (e.g., Cushing et al. 2008, Rice et al. 2010, Stephens et al. 2009, Liu et al. 2011). In some cases these {   few} parameters are not enough to adequately explain the spectra. Often times the fits are rather poor making any inference about these quantities and their corresponding uncertainties invalid (e.g., Czekala et al. 2015). This suggests that the physical assumptions made within the grid models are not adequate to describe the data and that the grid models are underutilizing the full information content of the data. 

To remedy these issues some modelers chose to add additional physical parameters to their models such as disequilibrium chemistry via an eddy diffusivity (Saumon et al. 2006; Stephens et al. 2009) and more sophisticated cloud parameterizations (Ackerman \& Marley 2001; Cushing et al. 2008). {   In these cases it is possible to get better fits but still with unphysical values for some parameters, suggesting that something else is missing}. Furthermore, often-unknown differing assumptions within the models from different groups, given the same sets of basic parameters, do not agree resulting in inconsistent interpretations of the same datasets (Figure 8, Patience et al. 2012). 

Motivated by these shortcomings, Line et al. (2014) \& (2015) ({   Line et al. 2015 referred to as ``Part 1" hereafter}) built upon the success of atmospheric retrieval methods applied to Earth (Rodgers 1976, Towmey 1996, Rodgers 2000, Crisp et al. 2004), solar system bodies (Conrath et al. 1998, Irwin et al. 2008, Fletcher et al. 2007, Greathouse et al. 2011), and exoplanet atmospheres (Lee et al. 2012; Line et al. 2012;13; Benneke \& Seager 2012; Barstow et al. 2013;14, Waldmann et al. 2015a,b) to develop a framework for retrieving the molecular abundances, vertical temperature structure, gravity, and other quantities. {   Such an approach is agnostic as to the detailed physical and chemical mechanisms governing the atmosphere; rather it allows the data to directly determine the fundamental quantities that impact the emergent spectrum, such as the opacities (governed by molecular abundances and cloud properties) and radiative source functions (governed by the thermal structure). It is from these diagnostic quantities that we can infer the properties of chemical and physical processes that are typically assumed in grid models. In short, self-consistent predictive grid models describe the state of the atmosphere with far fewer parameters (e.g., effective temperature and gravity) but incorporate more assumed physical processes that map those parameters onto the observable.  The retrieval approach requires many more parameters but fewer assumptions.}  

Part I leveraged the power of two benchmark brown dwarf-stellar systems (Gl570 and HD3651) for which we know the host star ages, metallicities, and a new dimension, carbon-to-oxygen ratios (C/O's).  The goal of Part I was to demonstrate the feasibility of our approach by retrieving these quantities from the companion brown dwarfs. Assuming co-evolution, the brown dwarfs should have the same ages, metallicities, and C/O's as the host star.    With low-resolution near-infrared SpeX data of Gl570D and HD3651b, {   Part I} was able to show that the retrieval method was indeed able to reproduce these quantities to within the errors, thus validating the approach.  {   We} also showed, for the first time, that one could robustly detect the presence of ammonia in low-resolution near-infrared  data, without resolving individual spectroscopic features. This work demonstrated the power of using a Bayesian approach to detect molecules that would have only otherwise been detected at higher resolution (e.g., Canty et al. 2015) or at longer wavelengths (e.g., Saumon et al. 2006; Hubeny \& Burrows 2007). Furthermore, {   we} were able to rule out the presence of optically significant clouds within the observable portion of the atmosphere of these objects.  Finally, via the retrieved temperature structures we were able to validate the radiative equilibrium assumption commonly made in self-consistent grid models.  

\begin{figure}
\includegraphics[width=0.5\textwidth, angle=0]{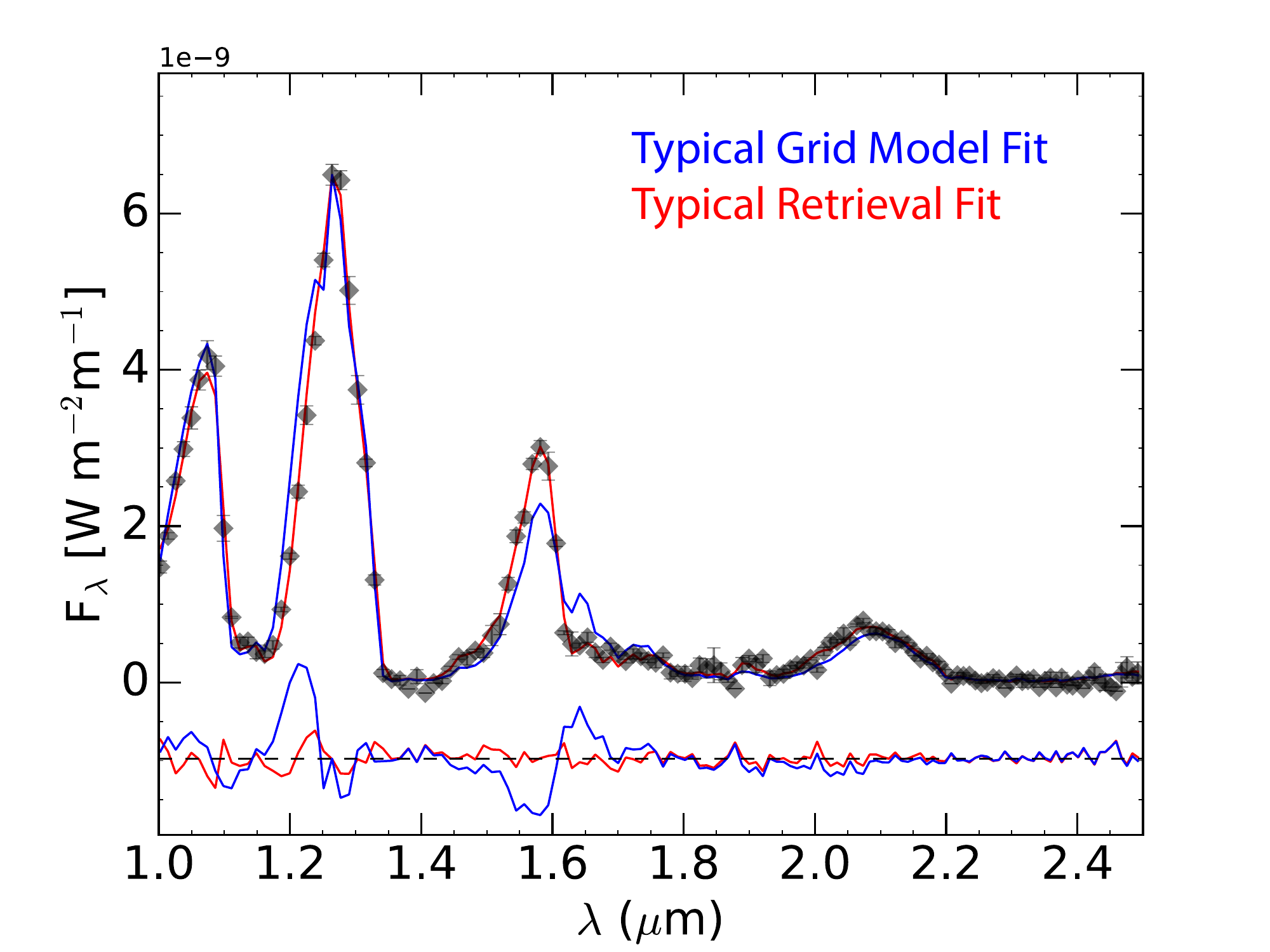}
     \caption{ \label{fig:figure0}  Comparison between retrieval fits and self-consistent grid model fits {   on the benchmark system, Gl570D (gray diamonds)}.  The fit residuals are shown at the bottom of the plot.  The grid model fit (in blue) is based on a standard set of assumptions (e.g., Saumon \& Marley 2008; Morley et al. 2012) involving 4 free parameters: the effective temperature, gravity, cloud sedimentation (Ackerman \& Marley 2001), and a flux scaling factor.  The nominal best fit grid model (effective temperature=900K, log-gravity=5., no cloud ) results in a chi-square per data point of 14.5, with 139 data points . With 4 free parameters, this results a Bayesian Information Criterion Value of 2035.2.  In contrast the retrieval model (red) with 27 free parameters produces a chi-square per data point of 5.0 and a BIC of 832.4. Despite the factor of $~7$ in the number of free parameters, the retrieval model is overwhelmingly favored with a $\Delta$BIC of $\sim$1200, which would be considered ``Very Strong" on the Jeffery's scale.  }
  \end{figure}

To illustrate the differences between a {   model fit obtained within a typical ``grid-model" framework and one within a``retrieval" framework} we compare a representative grid model best-fit (Morley et al. 2012) to a representative retrieval model (Part I) best fit in Figure \ref{fig:figure0} on the benchmark T-dwarf, Gl570D.   The grid models are based on Morley et al. (2012) with the spectra being computed on a grid as a function of 4 free parameters:  effective temperature, gravity,  clouds parameterized with a sedimentation efficiency (Ackerman \& Marley 2001), and a radius-to-distance scale factor (see Morley et al. 2012 for details).  Certainly, different grid models may have different parameters, and may fit either better or worse than the one chosen here. This is simply meant to be a representative example.   Note that the retrieval fit (red) is significantly better than the grid model fit (blue).  More quantitatively, the retrieval produces a chi-square per data-point of 5, where-as the best-fit grid model a chi-square per data-point of 14.5.  The model used in the retrieval from Part I included 27 free parameters, much more than the 4 used in the grid model.  Are the retrievals justified in using this large number of parameters? We turn to the Bayesian Information Criterion (BIC) which penalizes models with more free parameters. {   In this particular example, the retrieval fit is overwhelmingly favored, with a $\Delta$BIC difference of $\sim$1200 where a $\Delta$BIC$ >$10 is considered to be very strong evidence for the model producing the lower BIC value (e.g., Kass \& Raftery 1995), in this case, the fit resulting from the retrieval}.  One potential drawback of the retrieval models is, however, the possibility of unphysical solutions, {   though grid models are not always immune to this either.} As shown in Part I, the solutions obtained with our methodology are indeed chemically and physically plausible. We elaborate on this further in $\S$\ref{sec:chemistry}. 



In this manuscript we apply the Part I retrieval methodologies to a small sample (11) of late T-dwarfs (T7-T8) within the SpeX Prism Library (Burgasser et al. 2014) in order to identify diagnostic physical/compositional trends. {   The power of applying the same model to data taken with same instrument is that the results can be readily inter-compared, without worry of unknown model or data differences skewing inter-comparisons.}

We focus our analysis on late-T-dwarfs in this investigation for several reasons: firstly, {   they have deep spectral features due to strong water absorption and relatively cloud free atmospheres}, directly translating into a wide range of atmospheric pressures probed; This in turn should provide strong leverage on temperature-pressure profile constraints.  Secondly, the presumed relatively cloud-free atmospheres (e.g., Marley et al. 1996, Allard et al. 1996) mitigate significant temperature-cloud-gas abundance degeneracies. Thirdly, the cool temperatures ($\sim$600-800 K) of late-T-dwarfs are thermochemically predicted to result in large methane and water abundances (Lodders 2009). These species are both spectrally prominent with multiple bands across the near infrared, and are the dominant carbon and oxygen reservoirs, un-tainted by disequilibrium chemistry, allowing for robust metallicity and carbon-to-oxygen ratio determinations.  Finally, this temperature range encompasses the peak of the equilibrium temperatures spanned by most transiting exoplanets (Bathala et al. 2014).  Understanding the physical and chemical processes in these un-irradiated T-dwarf laboratories, with data of similar spectral quality as anticipated to come from the James Webb Space Telescope, will undoubtably provide a solid foundation and context with which we can base our planetary atmosphere inferences in the not too distant future.

 Our long-term goal is to extract molecular abundances, thermal structures, cloud properties, etc., for a large number of objects by working our way up (and down) the brown-dwarf spectral sequence, starting with these late T's then Y-dwarfs, followed by more cloudy mid-to-early T's and finally L-dwarfs.  By investigating ``simpler" {   (e.g., cloud free)} objects first, we can continue to identify the strengths and weaknesses of our model parameterization, making adjustments along the way.  

The manuscript is organized as follows: \S\ref{sec:Methods} summarizes the methodology outlined in Part I and identifies small changes.  \S\ref{sec:Sample} introduces our 11 T-dwarf sample. \S\ref{sec:Results} presents our retrieved quantities and the useful derived diagnostic quantities and trends. Finally,  \S\ref{sec:comp} compares our results to previous investigations and \S\ref{sec:Conclusions} summarizes our findings and discusses implications.

\section{Methods}\label{sec:Methods}
Here we summarize our methodologies developed in Part I. We couple a simple thermal emission radiative transfer forward model (Line et al. 2013a; 2014a,b; 2015)  with the powerful EMCEE sampler (Foreman-Mackey et al. 2014). The job of the forward model is to compute the observable given a set of parameters that describe the state of the atmosphere. In this case the observable is the disk-integrated top-of-atmosphere thermal flux with wavelength, {   a.k.a, the spectrum}. We neglect scattering and assume pure absorption {   This assumption will break down in the presence of strongly forward scattering clouds}.  The parameters that we use to describe the state of the atmosphere are the {   constant-with-altitude} volume mixing ratios for H$_2$O, CH$_4$, CO , CO$_2$, NH$_3$, H$_2$S, Na, and K, surface gravity, temperature-pressure (TP) profile, and the {   radius-to-distance dilution factor which effectively scales the top-of-atmosphere model flux to the appropriate distance and radius}.  We use the same novel TP-profile retrieval scheme described in Part I.   All other nuisance parameters from Part I are included (see Table \ref{tab:table1}).

\begin{table}
\centering
\caption{\label{tab:table1} Free parameters in the forward model (31 in total)}
\begin{tabular}{cc}
\hline
\hline
\cline{1-2}
Parameter\footnote{Part I had 27 free parameters for the cloud free model. The additional parameters here are the second smoothing prior hyper parameter (see footnote(b)) and the 3-cloud opacity profile parameters.}.  & Description    \\
\hline
log$f_{i}$ & log of the uniform-with-altitude\\
  &	 volume mixing ratios of H$_2$O, \\ 
   & CH$_4$, CO, CO$_2$, NH$_3$, H$_2$S, and alkali (Na+K)  \\
   log$g$ & log gravity [cms$^{-2}$]\\
   (R/D)$^2$ & radius-to-distance scaling (R$_{J}$/pc)\\
  T$_{j}$ & temperature at 15 pressure levels (K)\\
  $\Delta \lambda$ &  uncertainty in wavelength calibration (nm) \\
  $b$ & error bar inflation exponent\\
  	&	(Part 1, equation 3)\\
  $\gamma$, $\beta$\footnote{This is a new parameter not included in Part I. It adds flexibility to the inverse gamma hyper-prior distribution on $\gamma$. It could be considered a ``hyper-hyper-parameter". However, its influence was found to be negligible when comparing it to the nominal fixed value in Part I (see Table 2 in Part I).} & TP-profile smoothness hyperparameters \\
  				&(Part 1 Table 2 \& equation 5)\\
   $\kappa_{p_{0}}$, $P_{0}$, $\alpha$ & Cloud opacity profile parameters \\
   							 &  equation \ref{eq:cloud_eq} ($m^2/kg, bar, unitless$)\\

\hline
\end{tabular}
\end{table}
  
We discuss a few  {   points relevant to our modeling approach} not discussed in Part I.  Firstly, the statistics reviewer for Part I suggested we explore the impact of alternate priors, specifically the Dirichlet prior, which is suitable for compositional mixtures. Specifically we explored the Center-Log-Transform implementation (Aitchison et al. 1986; Benneke \& Seager 2012) on Gl570D (from Part I) and found no meaningful difference in the posteriors, at least for the set of 7 retrievable gases included in our current model. Therefore, for continuity with Part I we choose to continue to use uniform-in-log priors. 

Secondly, explore the impact of tabulated absorption cross-section resolution.  As in Part I, we use the absorption cross-section database described in Freedman et al. (2008,2014) and references there-in. Line broadening parameters are drawn from a variety of sources discussed within Freedman et al. (2008; 2014). This database contains pre-computed cross sections on a variable resolution wavenumber grid that samples the lines at 1/4 of their Voigt half widths on a pressure-temperature grid spanning 75 and 4000K and 300 - 1E-6 bars (see Freedman et al. 2008 for details of generation of the database). This resolution is fine enough to resolve the individual molecular lines over the wavelengths and environmental conditions we are interested in, and for all intents and purposes can be considered line-by-line.  

\begin{figure}
\includegraphics[width=0.5\textwidth, angle=0]{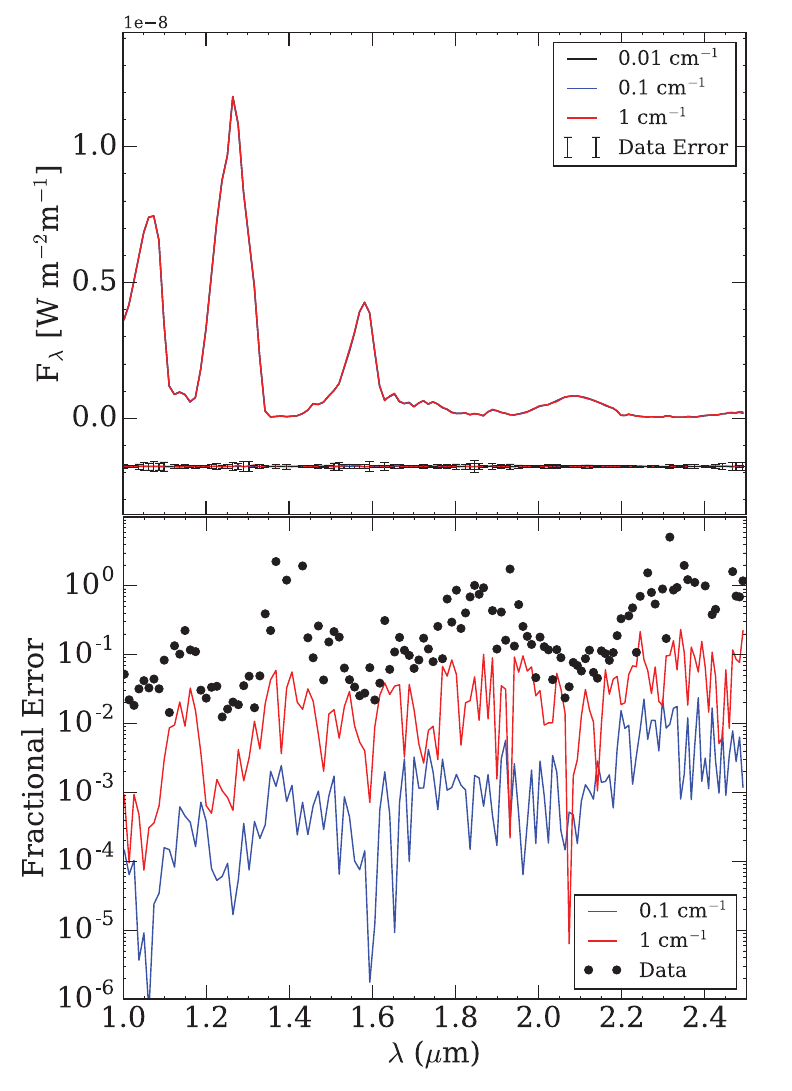}
     \caption{ \label{fig:figure1}  Validation of cross-section sampling method.  We compare a representative model spectra generated from cross-sections sampled from 0.01, 0.1, and 1 cm$^{-1}$ convolved and binned to the SpeX {   prism} instrumental resolution (R$\sim$80-120, depending on wavelength).  The top panel directly compares the binned model spectra and the differences between the 0.1 and 1 cm$^{-1}$ sampling and the line-by-line (0.01 cm$^{-1}$) to typical observational error bars. The bottom panel compares the relative (with respect to the 0.01 cm$^{-1}$ sampling) fractional error in the 0.1 and 1 cm$^{-1}$ sampled models to typical relative observational errors.  The 1 cm$^{-1}$ sampling errors are typically well below the relative observational errors, and is therefore a sufficiently high enough sampling resolution for the SpeX {   prism} instrumental resolution.  }
  \end{figure}

Of course we do not perform our retrievals at such fine resolution as that would result in optical depth computations for $\sim\times10^6$ lines--not computationally practical for a retrieval. Instead we use a cross section sampling method to reduce this number. We employ the method described in Sharp \& Burrows (2007) where we down sample the high-resolution cross-section database to 1 cm$^{-1}$ resolution by interpolating the line-by-line grid to the courser resolution grid. This has been suggested in Sharp \& Burrows (2007) to be a good balance between accuracy and efficiency.  Figure \ref{fig:figure1} shows a model spectrum convolved to a typical low-resolution SpeX observation (wavelength dependent resolving power, R$\sim$80-120) under 0.01, 0.1, and 1 cm$^{-1}$ sampling resolutions. We find that the residuals with respect to the line-by-line model (taken to be the 0.01 cm$^{-1}$ resolution) are minimal (typically an order of magnitude less relative uncertainty) relative to typical observational uncertainties, and therefore have no impact on our retrieved quantities. 

Part I used a simple gray slab, non-scattering cloud scheme and found its inclusion had virtually no impact on any of the other retrieved quantities for Gl570D and HD3651b. Because there are {   warmer T-dwarfs in our new sample, where clouds may potentially become more important},  we use a somewhat more complex cloud parameterization  (though it turns out not to matter) to account for possible vertical variation in the cloud profile. We use similar parameterization as Burrows, Sudarsky, \& Hubeny (2006) where the cloud opacity, $\kappa_{c}(P)$ is given by
\begin{equation}\label{eq:cloud_eq}
\kappa_{c}(P)=\kappa_{P_0}\left(\frac{P}{P_0}\right)^{\alpha}
\end{equation}
where $P_{0}$ is the cloud base pressure, $\kappa_{P_0}$ is the specific absorption coefficient (area/mass) at the base of the cloud, and $\alpha$ describes shape of the cloud profile ({   or how quickly the cloud opacity decreases with decreasing pressure from the base of the cloud)}.   Given the lack of evidence for clouds in the two objects explored in Part I, and the lack of detailed cloud-diagnosing spectroscopic features over the SpeX wavelengths, more complex parameterizations are unjustified (e.g., Morley et al. 2012).

\section{Late T-Sample}\label{sec:Sample}
As in Part I we draw from the SpeX Prism Library\footnote{http://pono.ucsd.edu/$\sim$adam/browndwarfs/\\spexprism/html/tdwarf.html}  (Burgasser et al. 2014) due to the continuous spectral coverage and instrumental uniformity across hundreds of objects. We selected 11 objects with known parallaxes that fell within the T7-T8 spectral classes and had near-infrared spectra in the original SpeX Prism Library as of 2015.  The normalized spectra are calibrated to flux units using the approach in Part I, but instead of using 2MASS photometry, we now use H-band photometry from the literature on the Mauna Kea Observatories (MKO) photometric system (Simons et al. 2002; Tokunaga et al. 2002 ).  We briefly describe the state of knowledge of each of the objects below {   focusing on metallicity, multiplicity, variability, and evolutionary properties} (of which is very heterogeneous).  Table \ref{tab:table2} provides the {   spectral types, the MKO H magnitude, parallaxes, and associated references} required for this study, and Table \ref{tab:table3} shows the published grid-model atmosphere fits of these objects spectra and the resulting physical parameters, typically gravity and effective temperature.  Derivations of physical parameters using evolutionary models, i.e., based on the object's bolometric luminosities and assumed ages, are not included.

\begin{table*}
\centering
\caption{\label{tab:table2} Summary of basic sample properties }
\begin{tabular}{llll}
\hline
\hline
\cline{1-2}
Object & SpT NIR                                        &       H$_{MKO}$ (mag)              &       $d_{\pi}$(pc)             \\
\hline
HD 3651B$^{1}$ & T7.5$^{9}$                                    & 16.68$\pm$0.04$^{9}$         &        11.06$\pm$0.04$^{17}$       \\
2MASS J00501994-3322402$^{2}$ &T7$^{10}$             & 16.04$\pm$0.10$^{11}$        & 	 10.57$\pm$0.27$^{13}$       \\
2MASSI J0415195-093506$^{3}$&T8$^{10}$             & 15.67$\pm$0.01$^{12}$         &  5.71$\pm$0.06$^{13}$             \\
2MASSI J0727182+171001$^{3}$ &T7$^{10}$            & 15.67$\pm$0.03$^{12}$         &  8.89$\pm$0.07$^{13}$            \\
2MASS J07290002-3954043&T8 pec$^{4}$      & 16.05$\pm$0.18$^{13}$          &  7.92$\pm$0.52$^{18}$             \\
2MASS J09393548-2448279$^{2}$&T8$^{10}$          & 15.96$\pm$0.09$^{14}$         &  5.34$\pm$0.13$^{19}$             \\
2MASS J11145133-2618235&T7.5$^{2}$        & 15.82$\pm$0.05$^{11}$          &  5.58$\pm$0.04$^{13}$      \\
2MASSI J1217110-031113$^{5}$&T7.5$^{10}$          & 15.98$\pm$0.03$^{15}$        & 11.01$\pm$0.27$^{20}$           \\
ULAS J141623.94+134836.3$^{6}$&T7.5$^{7,21}$        & 17.58$\pm$0.03$^{6}$        &  9.12$\pm$0.11$^{13}$        \\
Gliese 570D$^{8}$&T7.5$^{10}$                                     & 15.28$\pm$0.05$^{16}$        &  5.84$\pm$0.03$^{17}$       \\
2MASSI J1553022+153236AB$^{3}$&T7$^{10}$             & 15.76$\pm$0.03$^{12}$   &   13.32$\pm$0.11$^{13}$            \\
\hline
\end{tabular}
\tablerefs{
1 = \citet{2006MNRAS.373L..31M},
2 = \citet{2005AJ....130.2326T},
3 = \citet{2002ApJ...564..421B},
4 = \citet{2007AJ....134.1162L},
5= \citet{1999ApJ...522L..65B},
6= \citet{2010MNRAS.404.1952B},
7 = \citet{2010A&A...510L...8S},
8 = \citet{2000ApJ...531L..57B},
9 = \citet{2007ApJ...654..570L},
10 = \citet{2006ApJ...637.1067B},
11 = \citet{2010ApJ...710.1627L},
12 = \citet{2004AJ....127.3553K},
13 = \citet{2012ApJS..201...19D},
14 = \citet{2009ApJ...695.1517L},
15 = \citet{2002ApJ...564..452L},
16 = \citet{2010ApJ...710.1627L},
17 = \citet{2007A&A...474..653V},
18 = \citet{2012ApJ...752...56F},
19 = \citet{2008ApJ...689L..53B},
20 = \citet{2003AJ....126..975T},                                                             
21 = \citet{2010ApJ...725.1405B}}
\end{table*}

\noindent{   HD 3651B (T7.5)}: This is a wide (43\arcsec; 480~AU)
substellar companion to a K0V star, which also hosts a radial velocity
exoplanet ($M \sin{i} = 0.229\pm0.015$~\Mjup;
\citealp{2003ApJ...590.1081F}).  The stellar parameters and activity of
the primary star have been well-studied.   \citet{2006liu-hd3651b} summarized the available age data for
the primary based on chromospheric activity, X-ray emission,
gyrochronology, and stellar-isochrone fitting, adopting the last
method's result of 3--12~Gyr from \citet{2005ApJS..159..141V}.  Based on
this, \citet{2006liu-hd3651b} used evolutionary models and the T~dwarf's
bolometric luminosity to derive $\Teff = 810\pm30$~K and
$\logg = 5.3\pm0.2$.  More recent isochrone-based analyses give ages of
5.13~Gyr \citep{2007MNRAS.382.1516C}, 6.059~Gyr
\citep{2011A&A...532A..20F}, $6.9\pm2.8$~Gyr
\citep{2016A&A...585A...5B}, and $10\pm3.5$~Gyr \citep[][which includes
a direct stellar radius measurement from
interferometry]{2016A&A...586A..94L}.  \citet{mam08-ages} estimate an
age of 6.4--7.7~Gyr from chromospheric activity.   The primary star is well-established to have super-solar
metallicity, [Fe/H]=0.12--0.18, as summarized in Part I which also analyzes the C~and~O abundances.  
\noindent{   2MASS J00501994$-$3322402 (T7):}
\citet{2014ApJ...793...75R} report no variability in $J$~band, while
\citet{2014A&A...566A.111W} find a 10.8\% amplitude; re-analysis of the
latter's data by \citet{2014ApJ...797..120R} suggests the variability
detection is spurious.  \citet{2015ApJ...799..154M} robustly find mid-IR
[4.5] variability of 1.1\% with a periodicity of 1.55~hr, among the
shortest periods identified in optical/IR photometry of ultracool
dwarfs.  Among objects with mid-IR variablity, 2MASS~J0050$-$33 is
unique in that it is only variable at [4.5] and not at [3.6].

\noindent{   2MASSI J0415195$-$093506 (T8):} This well-studied object
is the T8 spectral standard for the \citet{2006ApJ...637.1067B} near-IR
classification scheme.  From near-IR and mid-IR data,
\citet{2007ApJ...656.1136S} inferred a likely super-solar metallicity
([Fe/H] = 0.0--0.3).  Based on IR colors and magnitudes,
\citet{2010ApJ...710.1627L} use models to infer that this object is
slightly cooler and probably more metal-rich than Gl~570D, in agreement
with the \citet{2007ApJ...656.1136S} spectroscopic analysis.
\citet{2010ApJ...722..682Y} analyze 2.5--5.0~\micron\ {\em AKARI}
spectra and find the non-equilibrium presence of CO and CO$_2$ bands
likely due to significant vertical mixing, in accord with the low
abundance of NH$_3$ in this object's mid-IR spectrum
\citep{2006ApJ...647..552S}.

\noindent{   2MASSI J0727182+171001 (T7):} This object is the T7
spectral standard for the \citet{2006ApJ...637.1067B} near-IR
classification scheme.  Based on a semi-empirical analysis of the
near-IR spectrum using model atmospheres from the Tucson group
\citep[e.g.][]{2005astro.ph..9066B}, \citet{2006ApJ...639.1095B} infer a
solar metallicity for the object.

\noindent{   2MASS J07290002$-$3954043 (T8 pec):}
\citet{2007AJ....134.1162L} suggest this object may be old, as its
anomalously brighter $Y$-band and fainter $K$-band peaks are suggestive
of suggestive of slightly low-metallicity and/or high gravity.

\noindent{   2MASS J09393548$-$2448279 (T8):} This object's near-IR
spectrum and luminosity are consistent with other objects of the same
spectral class, but its mid-IR spectrum and unusually bright mid-IR flux
point to a cooler object.  \citet{2006ApJ...639.1095B} suggest high
surface gravity and slightly subsolar metallicities given its anomalous
(broad) $Y$-band and (faint) $K$-band peaks compared to objects of
comparable spectral type.  \citet{2007ApJ...667..537L} reach the same
conclusion, and also estimate $[M/H]\approx-0.3$ from fitting the
near-IR spectrum.  Similarly, \citet{2009ApJ...695.1517L} compare the IR
colors with models and find a higher gravity and/or lower metallicity
than other late-T dwarfs.

Fitting the absolutely flux-calibrated near-IR and mid-IR data with
model atmospheres, \citet{2008ApJ...689L..53B} suggest the source is a
low-metallicity ($[M/H]\approx-0.3$) unresolved equal-mass binary, based
on the unrealistically large fitted radius.  Analyzing the same data,
\citet{2009ApJ...695.1517L} also favor a binary solution, either a pair
of similar 600~K dwarfs (whose temperatures would then be in discord
with the near-IR type of T8) or a dissimilar pair of 700~K and 500~K
dwarfs.  In either binary case, Leggett \etal\ infer slightly sub-solar
metallicity ($[M/H]\approx-0.3$--0.0).  The source is unresolved in
near-IR adaptive optics imaging down to 0.07\arcsec\ resolution (M. Liu,
priv. comm.).

Khandrika et al (2013) report a marginal detection of $K$-band
variability with large (0.3~mag) amplitude and indeterminate period.
However, \citet{2014ApJ...793...75R} do not find any $J$-band
variability down to $\approx$0.02~mag, nor do
\citet{2014A&A...566A.111W} report any $J$-band variability.

\noindent{   2MASS J11145133$-$2618235 (T7.5):} Like 2MASS~J0939$-$24,
\citet{2006ApJ...639.1095B} suggest high surface gravity and slightly
subsolar metallicity for this object based on its anomalously broad
$Y$-band and faint $K$-band peaks.  \citet{2007ApJ...667..537L} agree
with this assessment and also estimate $[M/H]\approx-0.3$ from fitting
the near-IR spectrum.

\noindent{   2MASSI J1217110$-$031113 (T7.5):}
\citet{2003ApJ...586..512B} report this has a faint 0.21\arcsec\
(2.3~AU) companion in their \HST\ optical imaging, but
\citet{2006ApJS..166..585B} conclude this was in fact an uncorrected
cosmic ray, based on a non-detection in their followup \HST\ near-IR
imaging.  However, given the 4-year gap between the optical and NIR
imaging, another possibility is that orbital motion moved the companion
too close to the primary to be observed in the followup epoch.  Assuming
a ratio of 1.08 between the semi-major axis and projected separation for
the moderate-discovery bias case of \citet{dupuy2011-eccentricity}, the
estimated orbital period would be 14--23~yr for a total system mass of
30--80~\Mjup, so significant orbital period over 4~years is plausible.
The $H$-band absolute magnitude (15.77 mag) is somewhat brighter than
typical for T7.5 objects (from Dupuy \& Liu 2012: $M_H = 16.3$~mag using
polynomial fit, 16.4~mag using averages of all T7.5's), though within
the RMS of these averages ($\approx$0.5 mag).  Thus, binarity remains a
possibility.

\citet{2006ApJ...639.1095B} note the brighter $K$-band peak suggests
reduced surface gravity compared to other late-T dwarfs, and their
spectral analysis infers solar metallicity.  \citet{2007ApJ...656.1136S}
analysis of the near-IR spectrum indicates enhanced metallicity, with
[Fe/H] close to +0.3, corroborated by the $H-[4.5]$ color analysis of
\citet{2010ApJ...710.1627L}.  The \citet{2007ApJ...656.1136S} \Spitzer\
mid-IR spectrum is not good enough to constrain the NH$_3$ abundance and
thus the degree of chemical disequilibrium, though the latter is
suggested by the relatively faint IRAC $[4.5]$ absolute magnitude in
comparison to the models.

\citet{2007ApJ...655.1079L} note this object's mid-IR colors (based on
\Lp, [3.6], and [4.5]) make it an outlier, not readily explained by
gravity, metallicity, or binarity effects.  These data may reflect
non-equilibrium effects, though the \citet{2007ApJ...656.1136S} modeling
does not really match the data, with possible $K_{zz}$ values spanning
an uncommonly large range of 10$^2$ to 10$^6$~cm$^2$~s$^{-1}$.

\citet{2014A&A...566A.111W} report possible $J$-band variabilty with
full amplitude of 4.2$\pm$1.1\%, while \citet{2014ApJ...793...75R}
report no $J$-band variability at the $\approx$0.03~mag level.
\citet{2014ApJ...793...75R} reanalyzed the strongly variable sources from
\citet{2014A&A...566A.111W} and found many were not in fact variable,
though her re-analysis did not include this object.

\noindent{   ULAS J141623.94+134836.3 (T7.5)} This is a very peculiar
object.  Its very blue $J-K$ and very red $H-[4.5]$ colors suggest low
metallicity, as does its unusual $J$-band peak shape and suppressed
$K$-band flux \citep{2010MNRAS.404.1952B}.
Fits to the near-IR spectrum by \citet{2010AJ....139.2448B} indicate
both high surface gravity and low metallicity ($[M/H/\le-0.3$).  (Note
that the spectra of \citealp{2010MNRAS.404.1952B} and
\citealp{2010AJ....139.2448B} significantly disagree about the
measurement of the red wing of the $J$-band methane absorption, an
unexplained anomaly.)  The object is a 9\arcsec\ separation (82~AU)
companion to the peculiar blue late-L dwarf SDSS J141624.08+134826.7
\citep{2010ApJ...710...45B, 2010AJ....139.1045S}, which also suggests
low metallicity though its kinematics are characteristics of a thin-disk
object.

\citet{2013AJ....145...71K} find possible variability with a 30-min
timescale and $\sim$11\% amplitude, and \citet{2014ApJ...793...75R} find
possible variability with $\approx$1\% amplitude.

\noindent{   2MASSI J1553022+153236AB (T7)} This is a 0.349\arcsec\
binary resolved with HST imagery and presumed to be a nearly equal mass T7/T6.5 pair, with $\Delta$F110W = 0.30~mag, estimated component spectral
types of T6.5 and T7, and an estimated mass ratio of 0.90$\pm$0.02
\citep{2006ApJS..166..585B}.

\noindent{   Gliese 570D (T7.5):} This is an extremely well-studied
object, given its brightness and the fact that it is a member of a
quadruple system that includes a M1V+M3V spectroscopic binary and a K4V
primary (GL~570A).  \citet{2006liu-hd3651b} estimate an age of 1--5~Gyr
based on the available age data at the time, with more recent estimates
of 1.36~Gyr from stellar activity \citep{2012AJ....143..135V},
6.7$_{-4.7}^{+4.8}$~Gyr from stellar isochrones
\citep{2011A&A...530A.138C}, and 3.7$\pm$0.6~Gyr from gyrochronology
\citep{2007ApJ...669.1167B}.  \citet{2006ApJ...647..552S} averaged
metallicity data from the recent literature to give [Fe/H] =
0.09$\pm$0.04, with more recent determinations of 0.31 from
\citet{2011A&A...530A.138C} and --0.05$\pm$0.17 from
{   Part I}.  

\begin{table*}
\centering
\caption{\label{tab:table3} Summary of spectroscopically derived physical parameters from the literature compared with our work.)}
\begin{tabular}{lcccccccl}
\hline
\hline
\cline{1-2}
 Object             &                  $\lambda\lambda$    &    \Teff    &      \logg    &    $[M/H]$    & log$(C/O)$\footnote{This is {\it not} relative to solar, but absolute. Solar C/O (0.55) would be a log$(C/O)$= -0.26 in this column.} &   $\log K_{zz}$  &    $R$   &   Ref  \\
               &                 (\micron)    &   (K)    &     (cgs)    &    (dex)    &   &   (cm$^2$ s$^{-1}$)  &    ($R_{J}$)   &    \\

\hline
HD 3651B                          & 1.0--2.1  & 790 $\pm$ 30  &5.0 $\pm$ 0.3&\nodata\footnote{Spectral fits assumed $[M/H]=0.12\pm0.04$, as derived
  from the primary star HD~3651A.} &  \nodata &  \nodata   & \nodata & 1\\
                                  & 0.7--2.5  &  820 -- 830   & 5.4 -- 5.5  & $\approx$0.2& \nodata  &   \nodata  &  \nodata   & 2\\
                                  & 1.15-2.25  &  800 -- 850   & 4.5 -- 4.8  & \nodata& \nodata  &   \nodata  &  0.996 -- 1.06   & 16\\
			         & 1--2.5  &  719$^{+19}_{-25}$   & 5.12$^{+0.1}_{-0.2}$  & 0.08$^{+0.05}_{-0.06}$ & -0.05$^{+0.09}_{-0.09}$   &  \nodata  &  1.10$^{+0.1}_{-0.07}$   & This Work\footnote{{   All of our retrieved quantities in this work are summarized with the median and 68\% confidence interval}} \\                                                                  
2MASS J00501994$-$3322402         & 1.0--2.1  & 960 -- 1000   & 4.8 -- 5.0  &      0      &  \nodata  & \nodata  &  \nodata   & 3 \\
                                			         & 1--2.5  &  815$^{+20}_{-27}$   & 5.09$^{+0.1}_{-0.2}$  & -0.06$^{+0.05}_{-0.06}$ & 0.06$^{+0.09}_{-0.1}$   &  \nodata  &  1.12$^{+0.12}_{-0.09}$   & This Work  \\                                                                  
                                                                     
2MASSI J0415195$-$093506          & 1.0--2.1  &  740 -- 760   & 4.9 -- 5.0  &      0      & \nodata  & \nodata  &  \nodata   & 3 \\
                                  & 0.7--14.5 &  725 -- 775   & 5.0 -- 5.4  & 0.0 -- 0.3  & \nodata  &    4     &  \nodata   & 4 \\
                                &1.147--1.347 & 947 $\pm$ 79  &4.3 $\pm$ 0.7& \nodata  &  \nodata   &  \nodata  &  \nodata   & 5 \\
                                  & 0.8--2.4  &  900, 1000    &   5.0       & \nodata  &  \nodata    &  \nodata  &  \nodata   & 6  \\                       
                                  & 2.5--5.0  &     800       &   4.5       &\nodata  &  \nodata    &  \nodata  &   1.14   & 7  \\
                                  & 2.5--5.0  &     900       &   4.5       &\nodata  &  \nodata    &  \nodata  &   0.66   & 8  \\
                                  & 1.0--5.0  &     700       &   4.5       &\nodata  &  \nodata    &  \nodata  &  \nodata   & 9  \\
                                  & 1.15-2.25  &  600 -- 800   & 4.0 -- 5.5  & \nodata& \nodata  &   \nodata  &  0.89 -- 1.33   & 16\\

			         & 1--2.5  &  680$^{+13}_{-18}$   & 5.04$^{+0.2}_{-0.2}$  & 0.05$^{+0.05}_{-0.07}$ & -0.01$^{+0.08}_{-0.1}$   &  \nodata  &  1.06$^{+0.05}_{-0.06}$   & This Work  \\                                                                  
                                                                                                     
2MASSI J0727182+171001            & 1.0--2.1  &  900 -- 940   & 4.8 -- 5.0  &      0      & \nodata  & \nodata  &  \nodata   & 3 \\
                                  & 0.8--2.4  &  1000, 1200   &   5.0       &\nodata  &  \nodata    &  \nodata  &  \nodata   & 6  \\                       
			         & 1--2.5  &  807$^{+17}_{-19}$   & 5.13$^{+0.1}_{-0.1}$  & 0.0$^{+0.04}_{-0.04}$ & -0.09$^{+0.07}_{-0.07}$   &  \nodata  &  1.12$^{+0.07}_{-0.06}$   & This Work  \\                                                                  
                                                                                                     
2MASS J07290002$-$3954043         & 1.0--2.1  &  740 -- 780   &     5.1     &\nodata  &  \nodata    &  \nodata  &  \nodata   & 10\\   
			         & 1--2.5  &  737$^{+21}_{-25}$   & 5.29$^{+0.1}_{-0.1}$  & -0.02$^{+0.05}_{-0.06}$ & -0.20$^{+0.09}_{-0.1}$   &  \nodata  &  0.95$^{+0.12}_{-0.10}$   & This Work  \\                                                                  
                                                                                                     
2MASS J09393548$-$2448279         & 1.0--2.1  & $\lesssim$700 &\nodata  &  \nodata    &  \nodata    &  \nodata  &  \nodata   & 3 \\
                                  & 0.7--2.5  &  725 -- 775   & 5.3 -- 5.4  &    $-$0.3   &\nodata  &  \nodata  &  \nodata   & 2 \\
                                  & 0.7--14.5 &     600       &     4.5     &    $-$0.3   & \nodata  &     4     &    1.26    & 11 \\
                                  & 0.7--14.5 &     600       &     5.0     &$-$0.3 -- 0.0& \nodata  &  \nodata  &    0.88    & 12\footnote{Assumes system is an unresolved equal-mass binary.} \\
                                  & 0.7--14.5 &     700       & 5.0 -- 5.3  &$-$0.3 -- 0.0&\nodata  &  \nodata  &0.78 -- 0.88& 12 (comp A)\footnote{Assumes system is an unresolved unequal-mass binary.} \\
                                  & 0.7--14.5 &     500       &     5.0     &$-$0.3 -- 0.0&\nodata  &  \nodata  &    0.88    & 12 (comp B)\footnote{Assumes system is an unresolved unequal-mass binary.} \\                                                                                                                                                                 
			         & 1--2.5  &  611$^{+17}_{-24}$   & 4.88$^{+0.2}_{-0.4}$  & -0.24$^{+0.07}_{-0.08}$ & -0.16$^{+0.13}_{-0.16}$   &  \nodata  &  1.22$^{+0.1}_{-0.09}$   & This Work  \\                                                                  
                                                                                                     
2MASS J11145133$-$2618235         & 1.0--2.1  & $\lesssim$700 &\nodata  &  \nodata    &   \nodata   &  \nodata  &  \nodata   & 3 \\
                                  & 0.7--2.5  &  725 -- 775   & 5.0 -- 5.3  &    $-$0.3   &\nodata  &  \nodata  &  \nodata   & 2 \\
			         & 1--2.5  &  678$^{+22}_{-22}$   & 5.13$^{+0.2}_{-0.3}$  & -0.06$^{+0.08}_{-0.08}$ & -0.31$^{+0.14}_{-0.14}$   &  \nodata  &  0.91$^{+0.09}_{-0.08}$   & This Work  \\                                                                  
                                                                                                     
2MASSI J1217110$-$031113          & 1.0--2.1  &  860 -- 880   & 4.7 -- 4.9  &      0      & \nodata  & \nodata  &  \nodata   & 3 \\
                                  & 0.7--14.5 &  850 -- 950   & 4.8 -- 5.4  & $\sim$0.3   &\nodata  &  2 -- 6   &  \nodata   & 4 \\
                                &1.147--1.347 & 922 $\pm$ 103 &4.8 $\pm$ 0.7&\nodata  &   \nodata   &  \nodata  &  \nodata   & 5 \\
			         & 1--2.5  &  726$^{+22}_{-25}$   & 4.74$^{+0.1}_{-0.2}$  & -0.12$^{+0.06}_{-0.07}$ & 0.04$^{+0.09}_{-0.11}$   &  \nodata  &  1.57$^{+0.11}_{-0.22}$   & This Work  \\                                                                  
                                                                                                     
ULAS J141623.94+134836.3          & 0.9--2.4  & 650 $\pm$ 60  &5.2 $\pm$ 0.4& $\le-0.3$   &\nodata  &     4     &  \nodata   & 13 \\                         
			         & 1--2.5  &  605$^{+29}_{-35}$   & 4.93$^{+0.4}_{-0.4}$  & -0.35$^{+0.10}_{-0.11}$ & -0.35$^{+0.20}_{-0.19}$   &  \nodata  &  0.8$^{+0.07}_{-0.06}$   & This Work  \\                                                                  
                                                                                                     
2MASSI J1553022+153236AB        &1.147--1.347 & 941 $\pm$ 138 &4.6 $\pm$ 0.6&\nodata  &  \nodata    &  \nodata  &  \nodata   & 5\footnote{Fit does not account for binarity.} \\
			         & 1--2.5  &  803$^{+16}_{-27}$   & 4.80$^{+0.1}_{-0.2}$  & -0.19$^{+0.04}_{-0.06}$ & -0.11$^{+0.09}_{-0.09}$   &  \nodata  &  1.59$^{+0.14}_{-0.09}$   & This Work  \\                                                                  
                                                                                                     
Gliese 570D                       & 1.0--2.1  &  780 -- 820   &   5.1       &      0      &\nodata  &  \nodata  &  \nodata   & 3 \\
                                  & 0.7--14.5 &  800 -- 820   & 5.1 -- 5.2  &      0      &\nodata  &6.2 $\pm$ 0.7&  \nodata & 14, 15 \\
                                &1.147--1.347 & 948 $\pm$ 53  &4.5 $\pm$ 0.5&\nodata  &  \nodata    &  \nodata  &  \nodata   & 5 \\
                                  & 0.8--2.4  &     900       &   5.0       &\nodata  &  \nodata    &  \nodata  &  \nodata   & 6  \\                       
                                  & 1.0--5.0  &     700       &   4.5       &\nodata  &  \nodata    &  \nodata  &  \nodata   & 9  \\    
                                  & 1.15-2.25  & 800 -- 900   & 4.5 -- 5.0  & \nodata& \nodata  &   \nodata  &  0.903 -- 1.06   & 16\\
                   
			         & 1--2.5  &  715$^{+20}_{-22}$   & 4.80$^{+0.3}_{-0.3}$  & -0.15$^{+0.07}_{-0.09}$ & -0.10$^{+0.13}_{-0.15}$   &  \nodata  &  1.14$^{+0.1}_{-0.09}$   & This Work  \\                                                                  

\hline
\end{tabular}
\tablerefs{
1 = \citet{2007ApJ...658..617B},
2 = \citet{2007ApJ...667..537L},
3 = \citet{2006ApJ...639.1095B},
4 = \citet{2007ApJ...656.1136S},
5 = \citet{2009A&A...501.1059D}, 
6 = \citet{2009A&A...503..639T},
7 = \citet{2010ApJ...722..682Y}, 
8 = \citet{2011ApJ...734...73T},
9 = \citet{2012ApJ...760..151S},
10 = \citet{2007AJ....134.1162L},
11=\citet{2008ApJ...689L..53B},
12 = \citet{2010ApJ...710.1627L},
13 = \citet{2010AJ....139.2448B},
14 = \citet{2006ApJ...647..552S},
15 = \citet{2009ApJ...695..844G} 
16 = \citet{2011ApJ...740..108L}} 

\end{table*}

\section{Results}\label{sec:Results}
\begin{figure*}
\includegraphics[width=1\textwidth, angle=0]{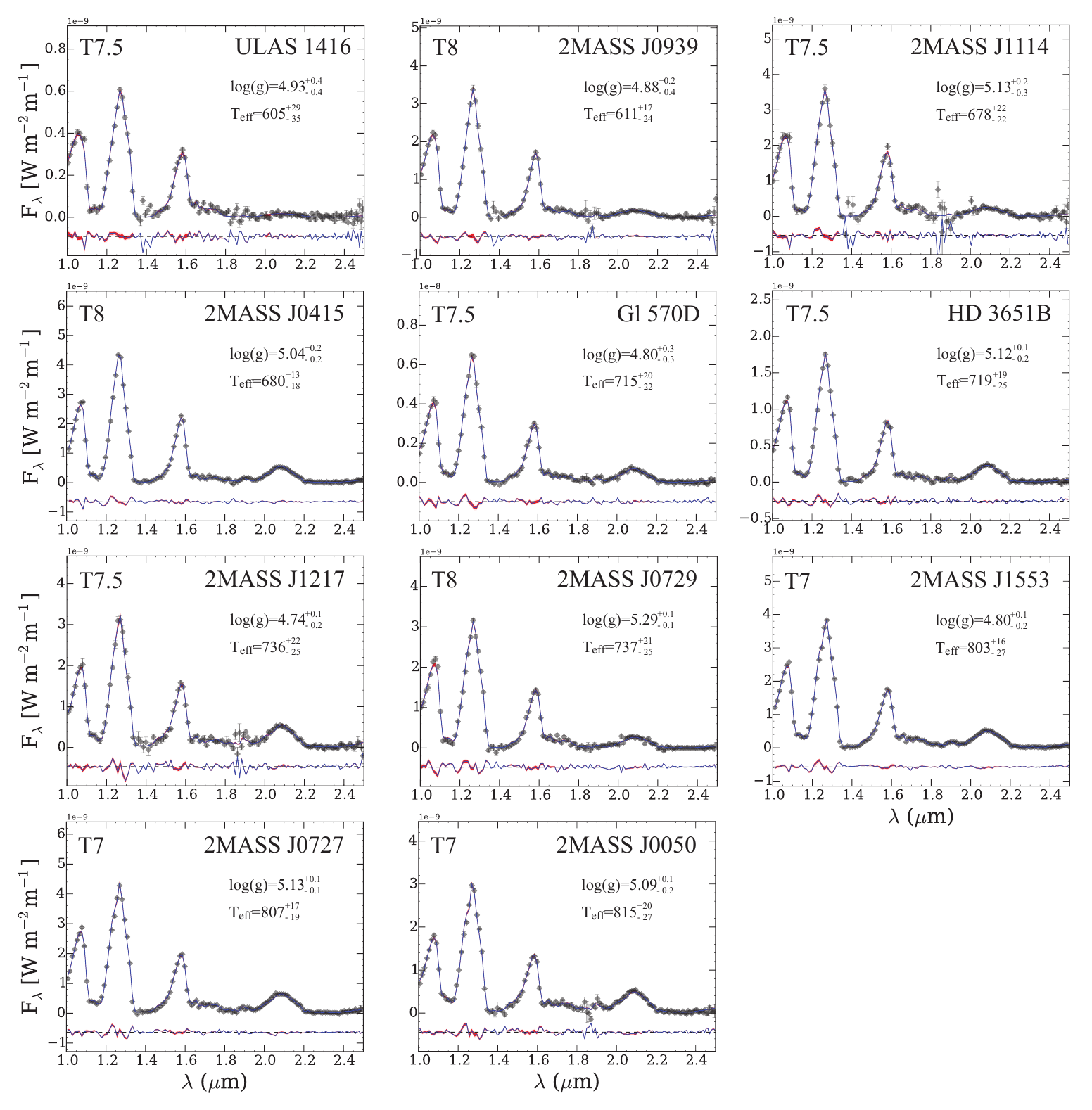}
     \caption{ \label{fig:figure2}  SpeX data along with their fits, sorted by effective temperature.  The data are indicated with the gray diamonds with error bars ({   typically equal to or smaller than the point size}).  The retrieval fits are summarized with a median (blue), 1- and 2-sigma confidence intervals (red, pink respectively) for spectra computed from 1000 random parameter vectors drawn from the posterior.   Note the small spread.  Residuals are shown below the spectra to illustrate the good quality of the fits.  The near-infrared spectral type is given in the upper left hand corner and the object name in the upper right.  The retrieved log$g$ (cgs) and derived effective temperature (K) and their uncertainties are also given.}
       \end{figure*}

\begin{figure*}
\includegraphics[width=1\textwidth, angle=0]{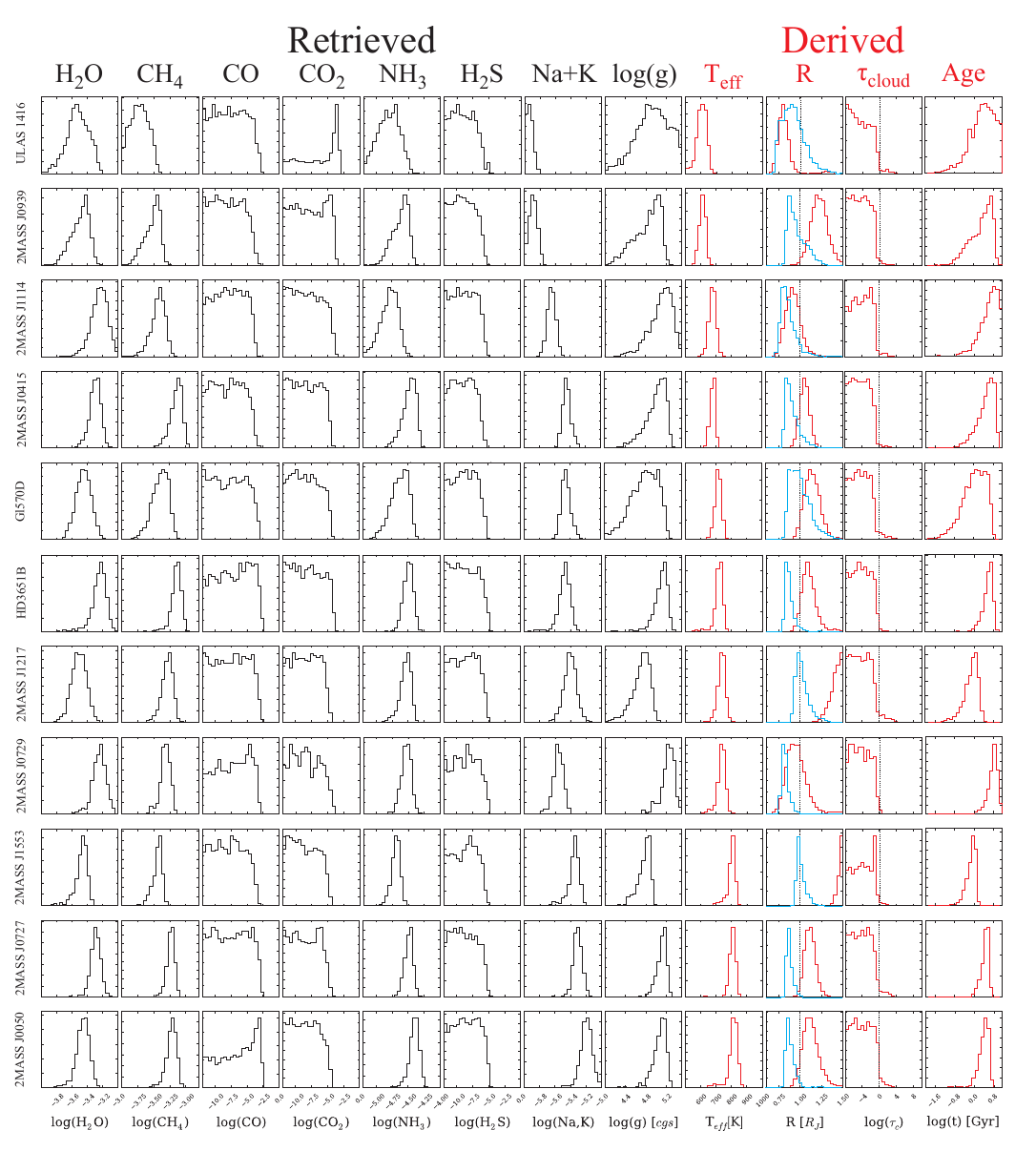}
     \caption{ \label{fig:figure3}  The marginalized posterior distributions of the relevant retrieved and subsequently derived quantities, sorted by effective temperature (low-to-high).  We show only the retrieved molecular abundances and gravity. All other parameters are either nuisance parameters (e.g., TP smoothing parameters, wavelength shifts, error bar inflation parameter) or summarized elsewhere (e.g., TP-profile).  The derived quantities (shown in red) are not directly retrieved from the data but are rather inferred/derived from quantities that are retrieved (see \S\ref{sec:Evo} for details). We also show for comparison (in blue) the evolutionary derived radius from the retrieved effective temperature and gravity using the relations in Burrows et al. (2001).  The dotted vertical lines represent 1 Jupiter radius.The cloud column optical depth ($\tau_{cloud}$) is derived by integrating the cloud opacity profile (obtained from the retrieved quantities in equation \ref{eq:cloud_eq}) over the region where a majority of the thermal emission contribution functions peak (see Figure \ref{fig:figure4}).    The dotted vertical lines denote an optical depth of unity.  Finally, the last column shows the evolutionary derived age from the effective temperature and gravity, again using the relations in Burrows et al. (2001). {   The full posteriors summarized via stair-step figures are available in the online supplemental material.} }
  \end{figure*}

Here we present the retrieval results for the spectra of 11 objects (Figure \ref{fig:figure2}). The relevant retrieved and subsequently derived quantities are summarized via marginalized posterior distributions (Figure \ref{fig:figure3}).   The retrieved temperature profiles are summarized in Figure \ref{fig:figure4}. We then present the molecular abundance results, their chemical plausibility, derived elemental abundances, resulting diagnostic trends, implications for cloud formation, and finally comment on the retrieved evolutionary related parameters (gravity, effective temperature, and radius).  {    All quoted and numerical values for retrieved/derived quantities refer to the median and 68\% confidence interval width derived from the posteriors of each parameter. Similarly, error bars in all figures represent the 68\% confidence interval about the median value for that quantity. The full posterior and correlations can be found in the supplementary documentation. There we also present python pickle files that contain the MCMC derived posterior and some summary data files.}.  

\subsection{Thermal Structures}\label{sec:TP}
Understanding atmospheric energy balance requires knowledge of the temperature-pressure profiles (TP-profile). The TP-profile of a sub-stellar atmosphere is largely governed by the opacity structure and internal heat flux, which is in turn dictated by the mass and age (e.g., Allard et al. 1996; Burrows 2001; Saumon \& Marley 2008). The two dominant energy transport mechanisms are radiation and convection.  Radiative energy transport tries to eliminate any vertical radiative flux divergence. Thermal structures that meet this criterion are said to be in radiative equilibrium.  At deep enough layers, where the mean opacities are large, radiative energy transport is no longer effective at carrying the heat flux.  At this point, convective energy transport dominates.  The location of this radiative convective boundary is predicted to occur at pressure levels deeper than $\sim$50 bars, or at atmospheric temperatures of $\sim$1500 K, for late T-dwarfs (Burrows et al. 2006).  Additional heating due enhanced opacity from cloud particulates can sometimes result in multiple convection zones separated by a radiative zone (Burrows et al. 2006).

Other non-traditional sources of heating/cooling that could impact the TP-profiles and/or their time dependence (e.g., Robinson \& Marley 2014) have been hypothesized.  Latent heat release due to cloud condensation can result in a shallower adiabat than predicted from pure gas phase chemistry alone (Tan \& Showman 2016), though the latent heat of typical L and T dwarf clouds are not particularly large (say, compared with water or ammonia clouds, which are important in cold solar system gas-giant atmospheres). Double diffusive convection resulting from vertical composition gradients can also result in shallower TP-profiles in the mid to deep atmospheres (Tremblin et al. 2015). Chromospheric heating can cause a more isothermal behavior in the upper ($P<\sim 1$ bar) atmospheres (Sorahana et al. 2014). Additional dynamical effects like gravity-wave breaking and horizontal dynamics can also cause perturbations to the TP-profile beyond the traditional assumption of 1D-radiative convective equilibrium. 

 {   Retrievals are in general agnostic to the physical mechanisms governing the TP-profile. Therefore,  empirically deriving the  TP-profiles through retrievals over a wide range of altitudes/pressures and identifying where these retrieved profiles differ from predictions made from typical self-consistent radiative-convective equilibrium profiles (which do not typically account for such additional energy sources) will allow us to diagnose where in the atmosphere ``non-standard" atmospheric energetic processes may play a role.  Such deviations, if at all present, motivate the addition of new physics to predictive forward grid-models.}. 

\begin{figure*}
\includegraphics[width=1\textwidth, angle=0]{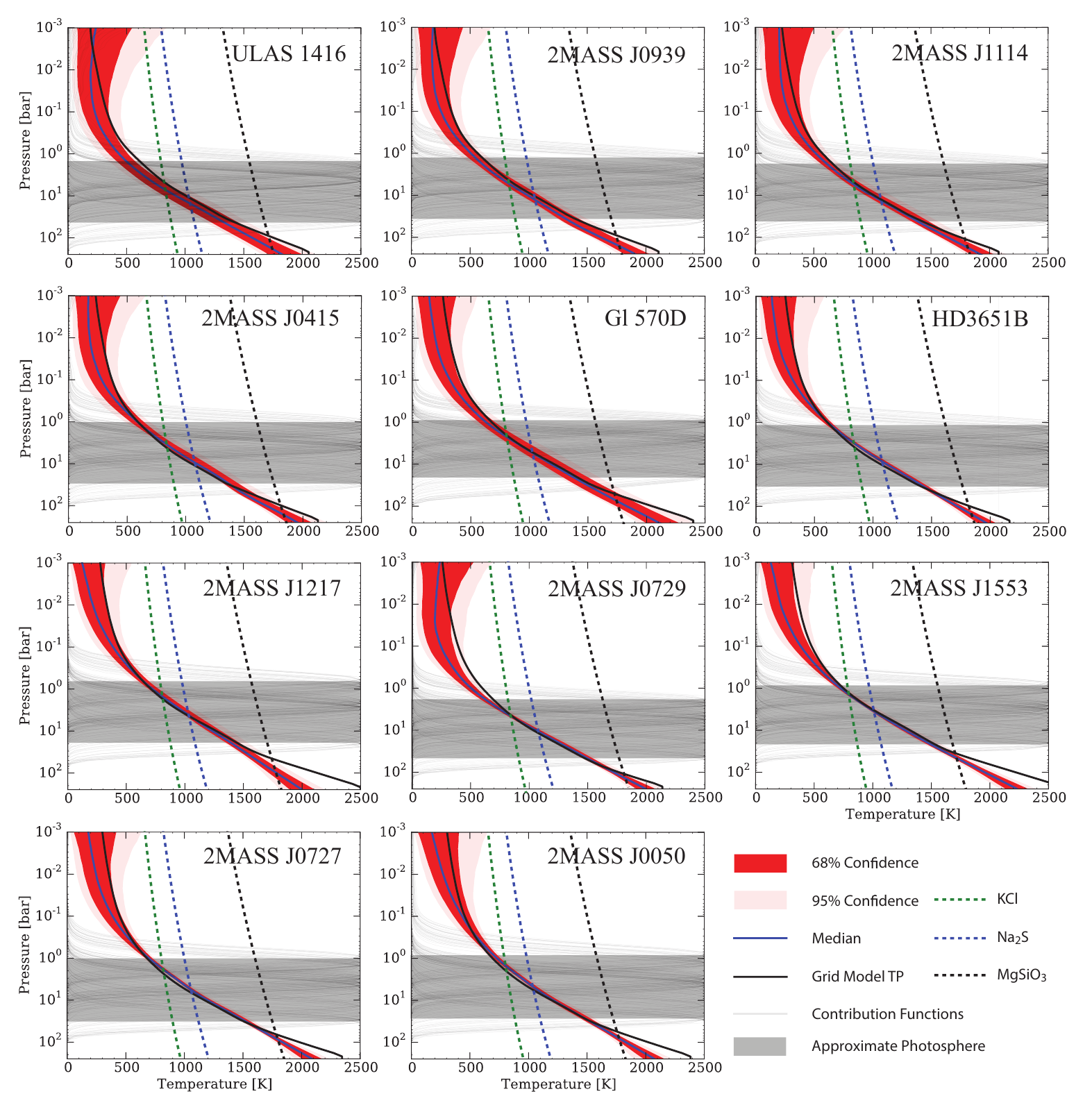}
     \caption{ \label{fig:figure4}  Retrieved temperature profiles sorted by effective temperature.  The median, 68\%, and 95\% confidence intervals are shown in blue, red, and pink, respectively.  The light gray curves are the thermal emission contribution functions for each SPeX wavelength for a representative fit. These represent the location of the so called  `photosphere'.  The black curve is a TP-profile from a self-consistent grid model (Saumon \& Marley 2008) corresponding to the median of the retrieved log($g$) and effective temperature.     The dashed curves are the equilibrium condensation curves for KCl, Na$_2$S, and MgSiO$_3$ from Morley et al. (2012) (and references therein), adjusted to the appropriate metallicity. The characteristic shape of the retrieved TP-profiles is in good agreement with self-consistent radiative-convective equilibrium predictions over the regions density probed by the contribution functions.  The divergence between the grid model and retrieved TP-profiles at the deepest pressure levels is largely due to the lack of spectral information probing those deeper levels. }
  \end{figure*}

{   Late T-dwarfs (and Y-dwarfs) offer the best chances of learning about the TP-profiles of brown dwarfs because of the large spectral dynamic range sculpted by strong water vapor absorption and their predominately cloud-free atmospheres. This high spectral dynamic range provides leverage on the TP-profile over a wide range of photospheric pressures.} Figure \ref{fig:figure4} shows the TP-profiles for the 11 T-dwarfs in our sample.   We also show a nominal profile {   from the Saumon \& Marley (2008) grid (interpolated to the median retrieved effective temperature and gravity)}.  Thermal emission contribution functions are shown to indicate the {   location of the photosphere\footnote{Because it is hard to formally define a photosphere for a brown dwarf given that the spectra probe a broad range of altitudes/pressures, we refer to photosphere here as the location where a bulk of the thermal emission contribution functions peak.}}, typically between $\sim$50 - 0.3 bar ($\sim$5 scale heights).  {   The emergent Y- and J- band spectra are shaped by the deepest, highest pressure layers, while the high opacity regions (dominated by water vapor) between these bands are shaped by the lowest pressure levels (highest altitudes)}.  The TP-profiles that overlap most with the thermal emission contribution functions are considered the most robust. Typical 1-sigma photospheric temperature uncertainties are  $\sim$30-100 K ({   Figure \ref{fig:figure4})} .  TP-profile information outside of those regions ($P>50$ bar and $P<0.3$ bar) are largely driven by our TP-profile parameterization (via the smoothing prior described in Part I) and should not be too heavily interpreted.  {   We also} show the equilibrium condensation curves for KCl, Na$_2$S, and MgSiO$_3$.  

In general, as we found for the two objects explored in Part I, the retrieved TP-profiles and predicted radiative-convective equilibrium grid models ({   Saumon \& Marley 2008}) agree quite well over the pressure levels densely probed by the spectra. Deviations begin to occur {   both above and below} where the contribution functions begin to wane. In Part I we suggested that the {   less steep} TP-profile gradient {   at pressures deeper than the} $\sim$50 bar level maybe indicative of the double-diffusive convection suggested by Tremblin et al. (2015). However,  there is very little contribution to the overall emission at these deeper levels due to the increasing opacity of the collision induced molecular hydrogen along with the alkali metal and water opacities at the J- and Y-band peaks.   At low pressure levels ($P<0.3$ bar), we typically find that the TP-profiles become more isothermal than predicted from radiative-convective equilibrium, possibly hinting at some sort of additional heating, but again temperature constraints are unreliable in this region.

\subsection{Composition \& Chemistry}\label{sec:comp}
We show, for the first time, the molecular abundances derived for an ensemble of T-dwarfs.  Molecular abundance determinations are diagnostic of the atmospheric chemical processes as well as the underlying elemental abundances.  We present the retrieved {   constant}-with-altitude molecular abundances for 7 gases: H$_2$O, CH$_4$, CO , CO$_2$, NH$_3$, H$_2$S, Na+K, where Na+K represents the combination of sodium and potassium, where we assume their ratio is assumed to be solar\footnote{We do this because the low-res SpeX data is really only sensitive to the red pressure broadened wings of these species, and to reduce the overall number of retrieved gases. Certainly the sodium-to-potassium ratio need not be solar, a quantity worth deriving from higher resolution spectra.}.    Figure \ref{fig:figure3} show the marginalized\footnote{{   Parameter correlations for these objects are quite similar to those shown in the stair-step plots for Gl570D and HD3651B in Part I. The main correlations are between the constrained species in gravity--e.g., a "metallicity vs. gravity" correlation, and some correlation amongst the gases themselves.  The full posteriors are shown in the supplementary material.}} gas posterior distributions for the 11 objects.  As in Part I, we find that H$_2$O, CH$_4$, NH$_3$, and the alkalis are the only species with bounded constraints {   on their abundances}. We can derive only upper limits {   on the abundances of} other species (CO, CO$_2$, H$_2$S), consistent with a non-detection of these gases (e.g., see the Bayes Factor analysis in Part 1). This is unsurprising given their anticipated low equilibrium abundance weighted cross-sections (Lodders et al. 2002 ) at near-infrared wavelengths.   {   In the following subsections we use our retrieved molecular abundances to, assess their chemical plausibility with respect to equilibrium chemistry, identify chemical processes driven trends with effective temperature, and determine elemental abundances (metallicities and carbon-to-oxygen ratios).}

\subsubsection{Comparison to Equilibrium Chemistry Predictions}\label{sec:chemistry}
\begin{figure*}
\includegraphics[width=1\textwidth, angle=0]{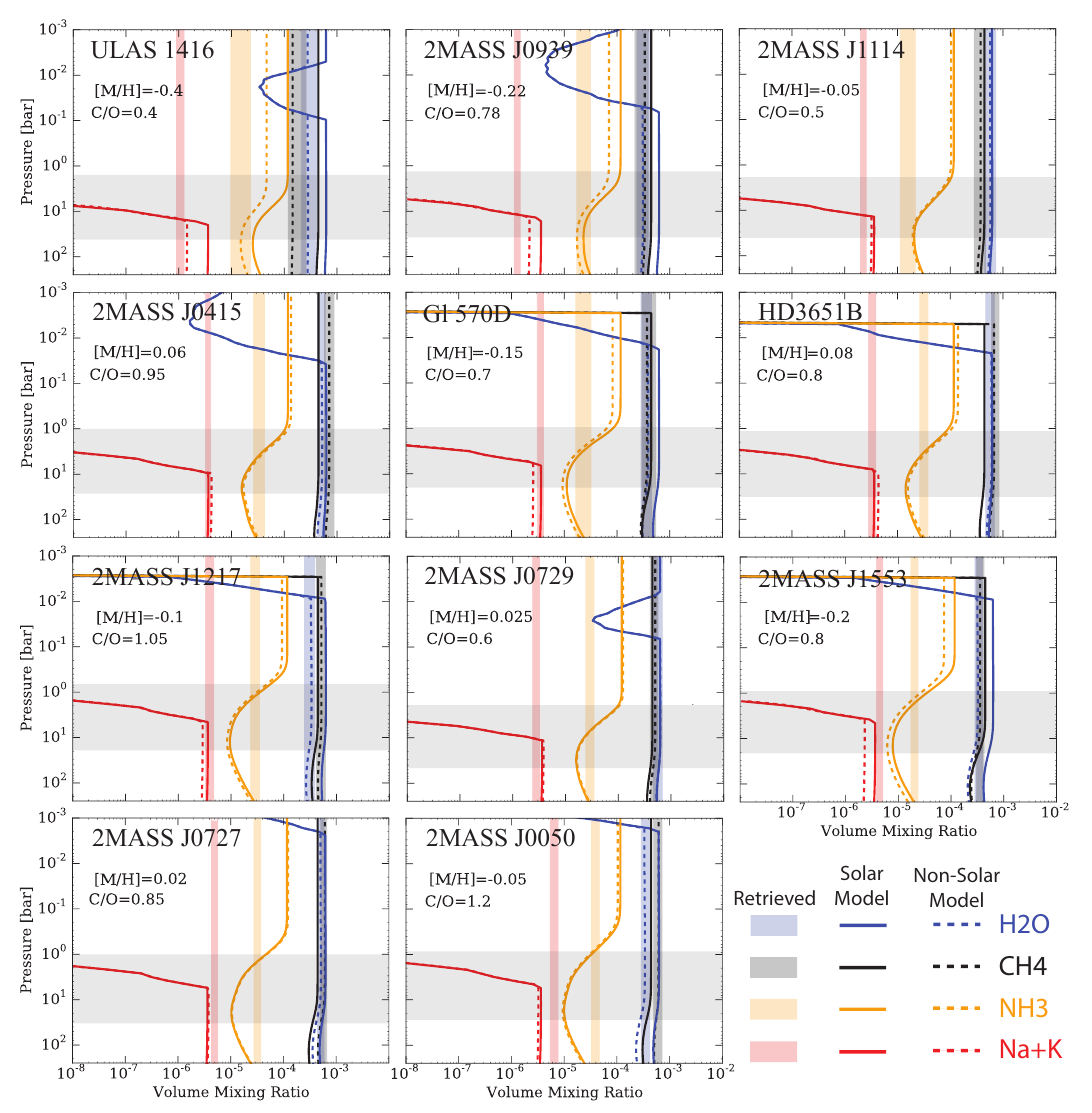}
     \caption{ \label{fig:figure5}  Comparison of retrieved abundances for well constrained species (Na/K, CH$_4$, H$_2$O, and NH$_3$) to thermochemical model predictions, sorted by effective temperature. The 1-sigma uniform-with-altitude mixing ratios retrieved from the data are indicated by the shaded rectangles. Thermochemical equilibrium molecular mixing ratios computed along the median TP-profile's assuming solar elemental abundances are shown as the solid curves. Finally, the dashed curves are the thermochemical equilibrium mixing ratios computed along the median TP-profiles assuming the metallicity, [M/H], and C/O values indicated in each panel. These quantities were adjusted to achieve a``by-eye" good match to the retrieved mixing ratios.   The metallicity here, [M/H], is defined as the log of all species heavier than hydrogen and helium relative to hydrogen, relative to that ratio for solar, e.g., [M/H]=0, and C/O=0.55 is solar where as [M/H]=1 would be 10$\times$ the solar metal abundance.  In general, the retrieved mixing ratios can be reproduced thermochemically given a proper tuning of the metallicity and C/O ratio, suggesting that the retrieved mixing ratios are thermochemically plausible. Note that we do not treat rainout here, rather adapt pure local thermochemical equilibrium for all $\sim$2000 species in the CEA/Thermobuild library (Gordon \& McBride 1996).   We do however, account for the depletion of oxygen due to enstatite rainout, which is predicted to remove 3.28 oxygen atoms per silicon atom (Burrows \& Sharp 1999).  The shaded gray region represents the approximate atmospheric pressures probed by the observations (see Figure \ref{fig:figure4})}
  \end{figure*}

Because we are retrieving each gas phase absorber individually, it is possible to obtain chemically implausible combinations (e.g., Heng \& Lyons 2016; Stevenson et al. 2014).  It is therefore prudent to check the plausibility of our results against a chemical model. For simplicity, we compare our results to predictions from a thermochemical model (Chemical Equilibrium with Applications Code 2, Gordon \& McBride 1996, see Part I for implementation details and relevant references) in Figure \ref{fig:figure5}.  Briefly, the model computes the thermochemical equilibrium molecular ratios at a given T-P pair and set of elemental abundances (Lodders \& Fegley 2002). The model accounts for both gas phase and condensed phase species in local thermochemical equilibrium, but does not take into account the rain-out paradigm whereby elemental species are depleted due to their sequestration into condensates that form large droplets and ``rain" out into the deeper unobservable atmosphere (e.g., Fegley \& Lodders 1994; Marley \& Robinson 2015).  We do account for the influence of the main rain out process (enstatite and forsterite condensation, Fegley \& Lodders 1994) on the elemental abundances (and subsequent equilibrium molecular abundances) by manually removing 3.28 oxygen atoms per silicon atom (Sharp \& Burrows 1999).

As in Part I, {   with the retrieved TP profiles} we can {   post}-compute the expected {   thermochemical} equilibrium molecular abundances along the median of the retrieved TP-profiles for each object (from Figure \ref{fig:figure3}). We compare the retrieved molecular abundances for only the well constrained species (shaded regions in Figure \ref{fig:figure5}) to the {   model} thermochemical equilibrium composition computed {   using} solar elemental ratios (solid curves in Figure \ref{fig:figure5}).  In general we find that the retrieved mixing ratios qualitatively agree with what is expected from thermochemical equilibrium at these temperatures:  CH$_4$ and H$_2$O are the most abundant species followed by NH$_3$ and then the alkalis. We emphasize that both our retrieved and thermochemically calculated mixing ratios are really only relevant over the region of atmosphere probed by the spectra ($\sim$30 - 0.3 bar, Figure \ref{fig:figure3}). Upon more careful inspection we find some inconsistencies between the thermochemical predictions at solar composition and the retrieved mixing ratios.  We then adjust the metallicity ([M/H])\footnote{``M" is the sum of all elements other than H and He. In this work, M is dominated by O, C, N, and Na+K.} of the thermochemical model by rescaling all elements heavier than H and He and the carbon-to-oxygen ratio (C/O) to by-hand ``fit" the thermochemical model (dashed curves in Figure \ref{fig:figure5} corresponding to the inset values) to the retrieved mixing ratios, as was done in Part I.   This is by no means a rigorous fit {   to the retrieved mixing ratios} (see \S\ref{sec:element_abund} for a more quantitative analysis) but is {   used only to show} that we can obtain consistency between the thermochemical prediction and the retrieved mixing ratios by adjusting these two quantities, suggesting that we have retrieved {\it chemically plausible} molecular abundance combinations.   

Finally, we find some inconsistencies between predicted and retrieved alkali abundances. This could be due to the {   difficulties in determining an appropriate description for the} pressure-broadened wings of the alkali lines (Sharp \& Burrows 2007; Allard et al. 2009) and/or the details of condensate rainout chemistry (which our model does not account for). {   Also, brown dwarf abundance determinations typically assume that elemental metal ratios are solar, i.e., if we increase the metallicity, all metal abundances increase by that same factor. However, this need not be the case within the context of galactic chemical evolution. For instance, Figure 16 of Hinkel et al. (2014) shows a $\sim0.8$ dex spread in the potassium-to-iron ratio for stars in the solar neighborhood. Such a spread could also imprint itself onto the brown dwarf population.   } 

\subsubsection{Retrieved Species Trends \& Possible Direct Evidence of Condensate Sequence}\label{sec:trends}
\begin{figure*}
\includegraphics[width=1\textwidth, angle=0]{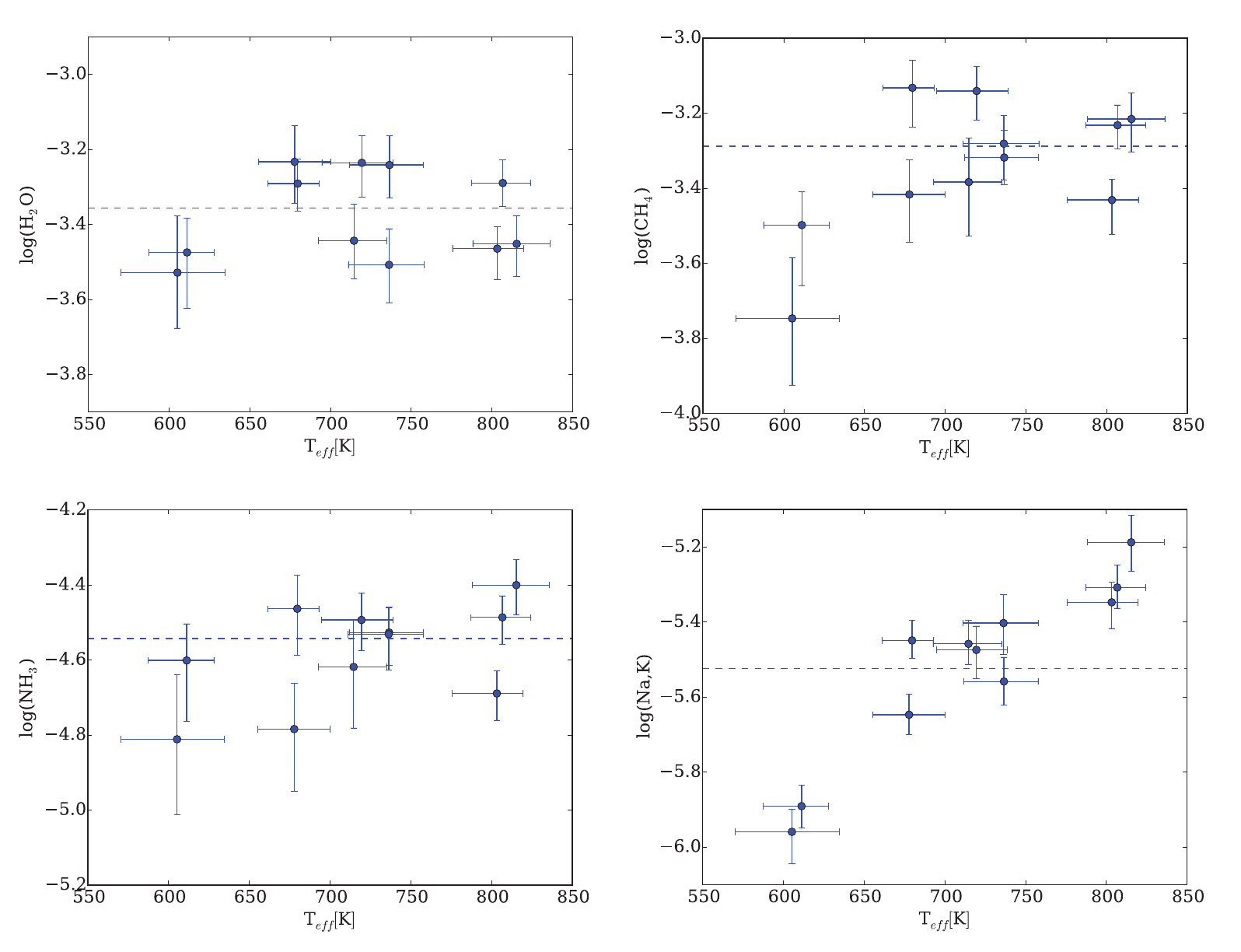}
     \caption{ \label{fig:figure7}  Molecular abundance trends with effective temperature for the well constrained species. The y-axis scale for each species spans 1 order of magnitude.  The horizontal dashed line in each panel is the weighted mean for each molecule, or equivalently, the best fit ``no-trend" constant. To compute the statistical significance of any trends, we rather determine how well we can reject the null (no trend) hypothesis. We can reject the null hypothesis at 1.0, 2.9, 0.9, and 26.9 $\sigma$ for water, methane, ammonia, and the alkalis, respectively. The alkali metals are the only species for which we can confidently reject the no trend hypothesis, all other trends are reasonably consistent with a horizontal line.    }
  \end{figure*}

An advantage of retrieval methods over grid models is that we can {\it directly} {   constrain} the molecular abundances rather than assume/compute them. {   From trends in those abundances (e.g., with temperature, gravity, etc.) we can infer physical/chemical processes}.  As an initial look, we focus on molecular abundance trends with effective temperature (Figure \ref{fig:figure7}). {   Chemical processes are predicted to depend strongly on temperature (e.g,  Burrows \& Sharp 1999, Lodders \& Fegley 2002); we would therefore expect some molecular abundances to vary with effective temperature}.    Specifically, we focus on trends associated {   with species that} are well constrained which includes H$_2$O, CH$_4$, NH$_3$, and the alkali metals. The other species (CO, CO$_2$, and H$_2$S), while chemically interesting,  do not have the {   retrieved abundance} constraints required to identify diagnostic trends.   

 We would predict water and methane to have little or no trend with temperature over our sample  ($\sim$ 600-850 K)  as they are thermochemically constant (e.g., Burrows \& Sharp 1999 Figure 5) and largely unaffected by disequilibrium processes (Saumon et al. 2006) at these temperatures. We indeed do not find any statistically robust trends of water and methane with effective temperature, as both are consistent with the null (flat line not rejected to $>$ 3$\sigma$) hypothesis (Figure \ref{fig:figure7}). 
 
 Ammonia (Figure \ref{fig:figure7}, bottom left), on the other hand, is thermochemically predicted to decrease by nearly an order of magnitude over this effective temperature range at the photospheric pressure levels (Burrows \& Sharp 1999).  Given the uncertainties, we should be able to identify such a trend if it existed; the lack of trend is interesting. Perhaps the lack of trend is speaking to perturbations due to disequilibrium chemistry dredging up ammonia from these deeper layers, {   though further work is necessary to robustly connect disequilibrium chemistry to the observed lack of trend}. {   Deriving the ammonia abundance for more objects over this temperature range should help refute or identify any trend if it existed.}
 
 
 A more striking possible trend is that of the alkali metal abundances with temperature (Figure \ref{fig:figure7}, bottom right),  with the null hypothesis being rejected at $>$25$\sigma$.  Over our effective temperature range, the alkali metal abundance varies by nearly an order of magnitude.  We attribute this trend to the sequestration of Na and K into salt sulfide (Na$_2$S) and potassium chloride (KCl) condensates (Lodders 1999;  Leggett et al. 2012; 2014; Liu et al. 2012; Schneider et al. 2015)  (dominated by the KCl condensation as K dominates the alkali absorption on the blue edge of the Y-band by a few to $\sim$1 order magnitude over Na at solar Na/K ratios)\footnote{while Na is predicted to be $\sim$20 times more abundant than K, the absorption cross section of K is $\sim$100 greater at shorter wavelength resulting in K dominating the opacity by nearly an order of magnitude}.  As these objects cool, the TP-profile-condensate intersection point {   ($P_{cond}$)} moves deeper into the atmosphere (as seen in Figure \ref{fig:figure4} where the dashed curves intersect the blue solid blue curve).   Gas phase sodium and potassium only exist in appreciable quantities {   at pressure levels {\it deeper} than $P_{cond}$}.  {   In cooler atmospheres, $P_{cond}$ occurs at higher pressures deeper in the atmosphere resulting in less alkali gas and more condensate over the portion of atmosphere probed by the spectra}.  
  
\begin{figure*}
\includegraphics[width=1\textwidth, angle=0]{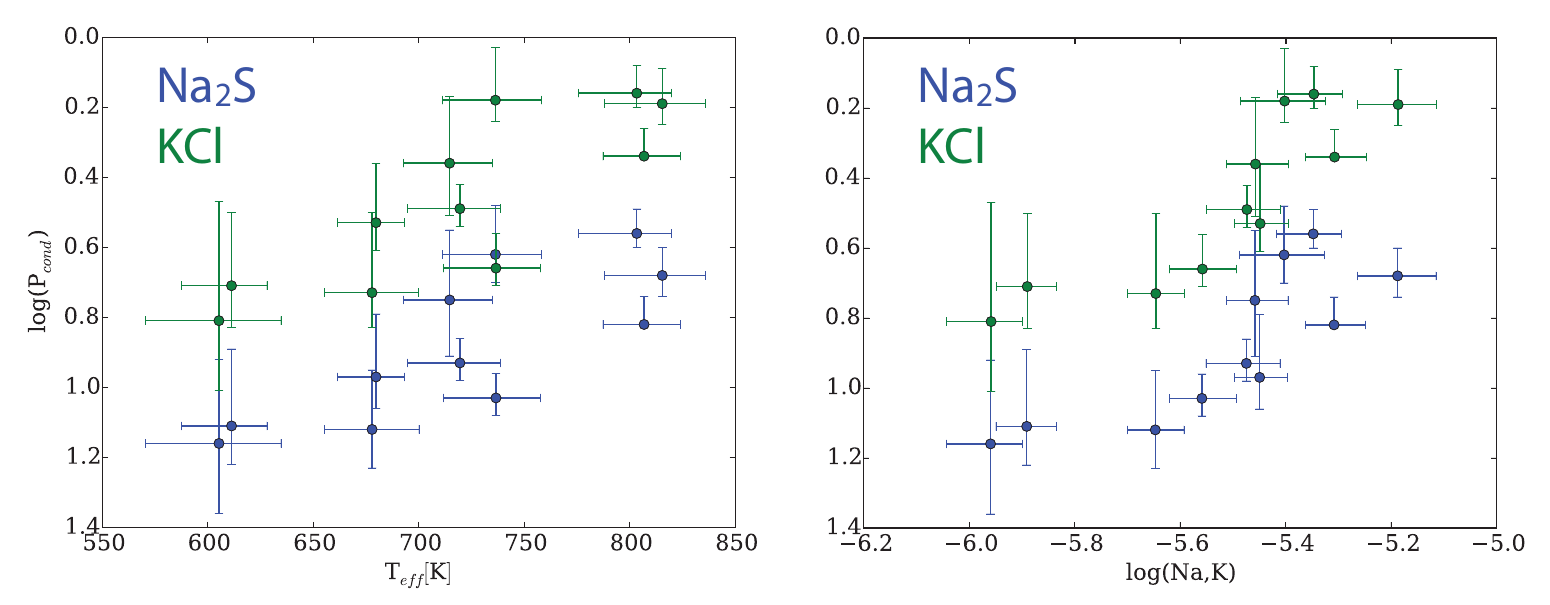}
     \caption{ \label{fig:figure8} Comparison of TP- condensate intersection pressure ($P_{cond}$) with effective temperature (left) and alkali metal abundance (right).  Hotter effective temperatures correspond to overall hotter TP-profiles (from Figure \ref{fig:figure4}) causing the intersection point to move towards lower pressures (higher altitudes).  As the TP-condensate intersection point moves towards lower pressures,  a greater integrated column abundance of the gas phase alkali metals is allowed (dominated by the K opacity), suggesting that the alkali trend is consistent with alkali rainout (see text).    }
  \end{figure*}

{   In Figure \ref{fig:figure8} we show how the $P_{cond}$ for Na$_2$S and KCl changes with effective temperature (left panel) and how the gas phase alkali abundances in turn depend on $P_{cond}$ (right panel) .  As expected, as cooler temperatures occur at higher pressures (deeper atmosphere) for cooler objects, the condensates are in turn expected to form at higher pressures (deeper atmosphere), thus resulting in a reduced gas phase column abundance of alkali metals, in turn resulting in the observed increasing trend in the alkali abundances with decreasing $P_{cond}$. }
 
 There have long been two approaches of how condensate chemistry should be treated in brown dwarf self-consistent atmospheric forward models. With the pure equilibrium approach, condensates are considered to continue to be in chemical equilibrium with their surrounding gas, even after condensation (for example Burrows \& Sharp 1999; Allard et al. 1999). With the second approach, commonly followed in solar system chemistry (Fegley \& Lodders 1994), condensed species `rain out' of the atmosphere and are no longer are available to react with surrounding gas at lower effective temperatures.
 
With the pure equilibrium approach gaseous Na and K react with previously condensed aluminum and silicate compounds to form feldspar minerals such as sanadine and albite (see Burrows et al. 2001 for a discussion).  {   The formation of these condensates would result in a depletion of the gas phase alkali abundances at temperatures between 1200-1400 K (Figure 13 of Burrows et al. 2001)}.  Meanwhile with the rainout approach the aluminum oxides and silicates rain out of the atmosphere after they condense and the feldspars do not form. Instead the gaseous Na and K form chloride salts and {   only begin to deplete the gas phase alkali's below effective temperatures of $\sim$1000K (Figure 12 of Burrows et al. 2001; Lodders 1999)}. {   A number of indirect lines of evidence (Marley et al. 2002, Morley et al. 2012; Schneider et al. 2015), have supported the rainout paradigm but a conclusive measurement of the Na and K depletion over the expected temperature range has been somewhat lacking.  With the decrease in the retrieved alkali abundances with decreasing effective temperature, and their correlation with the condensate base pressure, at effective temperatures below 1000K (otherwise we would detect no gas phase alkali's), we have shown that the rainout paradigm is indeed correct. }
 

\subsubsection{Metallicity \& Carbon-to-Oxygen Ratio Determination}\label{sec:element_abund}
The elemental abundances of a star or substellar object are diagnostic of its formation history and evolution as well as atmospheric chemical and physical processes (Bond et al. 2010; Burrows 2001; Marley \& Robinson 2015). Detailed stellar/substellar {   metal} abundances are also critical to the understanding of galactic chemical evolution (Timmes et al. 1995; McWilliam 1997; Venn et al. 2004).  
There has been a push to measure the elemental abundances present in stars in order to inform planet formation theories (e.g., Petigura \& Marcy 2011; Teske et al. 2014; Fortney 2012).  Much work has gone into determining stellar abundances, but it is by no means a solved problem as different methods and stellar parameter assumptions can result in different elemental abundance determinations (e.g., Hinkel et al. 2016).  

While {   metallicity measurements for solar-type} stars have received much attention, little work has been done in determining brown dwarf elemental abundances beyond their metallicities (e.g., Cushing et al. 2008; Stephens et al. 2009). Planets have been hypothesized (Payne \& Lodato 2007), and some detected (Han et al. 2013), around brown dwarfs.  Brown dwarfs will eventually provide a unique opportunity to detect/explore habitable planets via the transit method due to their favorable separation and radius ratios (Belu et al. 2013). We would therefore expect that in the not too distant future, brown dwarf elemental abundances determinations will become increasingly more important {   for planet formation theories, especially in light of the recently discovered TRAPPIST-1 planetary system (Gillon et al. 2016; 2017), hosted by a late M-dwarf.}  

The challenging aspect of elemental abundance determinations in cool objects is the presence of molecular species and condensates. Molecular species and condensates sequester the elements that would otherwise be available to form atomic absorption lines. Furthermore, broad molecular bands obscure the precious continuum so often required in classic stellar elemental abundance analysis (e.g., Bean et al. 2006). Therefore, novel methods need to be employed in order to translate the observed molecular absorption into the intrinsic elemental abundances.  Even then, molecular abundances can be perturbed by atmospheric processes like vertical mixing and rainout condensate chemistry. These processes need to be taken into account when inferring elemental abundances from molecular absorption. 

We derive the elemental abundances from the molecular abundances using ``chemical retrieval-on-retrieval" method described in Line et al. (2016) with the thermochemical equilibrium chemistry model described in $\S$\ref{sec:chemistry}.  We use only CH$_4$ and H$_2$O as the metallicity and C/O diagnostics since their mixing ratios are not strongly influenced by chemical quenching at these temperatures (in contrast to say, NH$_3$ and CO\footnote{{   These molecules, and all others, other than water and methane, are excluded from our elemental abundance analysis}}).  

The basic method is to fit the retrieved water and methane mixing ratios with the thermochemical equilibrium model with the free parameters being the C/O and metallicity ({   scaling to all elements heavier than H/He using the Lodders 2009 abundances}) using EMCEE.  The exact ``data" being fit here are the water and methane histograms from Figure \ref{fig:figure3}. At each MCMC step a TP-profile is randomly drawn from the retrieval posterior in order to propagate the TP-profile uncertainty into the C/O-metallicity results.  Given the randomly selected TP-profile, the C/O, and metallicity for that MCMC step, the thermochemical abundances of methane and water are computed at a representative photospheric pressure (3 bars\footnote{We are relatively insensitive to this value as the water and methane abundance profiles are fairly constant with temperature and pressure}, and the corresponding temperature at that pressure level), and evaluate their probability of occurrence from the methane and water histograms. These two probabilities are then multiplied and logged to compute the equivalent of ``chi-square".  Note that we could have a-priori fit the water and methane histograms with a Gaussian ahead of time to derive a ``data point" with an error bar of which we could then use the chi-square formula. However, the ``fit the histogram" approach is more rigorous because it can account for any arbitrarily shaped distribution (see Line et al. 2016 for details) and is fundamentally what minimizing chi-square tries to do anyway.

We apply this procedure under the two different condensate rainout paradigms. The first assumes no rainout, e.g., pure equilibrium. All condensates remain in the atmospheric layer at which the formed. The second is the silicate\footnote{We only account for silicate rainout as it is really the only set of condensates that impact the oxygen abundance. The alkali rainout discussed in \S\ref{sec:trends} won't impact the metallicities and C/O ratios derived from water and methane.} rainout paradigm whereby we remove 3.28 oxygen atoms per silicon atom (as was done in $\S$\ref{sec:chemistry})\footnote{Rainout likely happens for all condensates. Here we are only exploring the consequence of enstatite/forsterite rainout on the oxygen abundance}. Silicate condensation, and subsequent rainout, occur at the deepest layers inaccessible to the spectra where the TP-profiles intersect the silicate condensate curves.   Essentially, we are exploring the effect that different condensation assumptions can have on the derived elemental abundances. Unseen processes occurring at pressure levels {   above or below} where the spectra probe can impact the molecular abundances that we retrieve. With our two possible rainout assumptions {   are attempting to} account for these processes.

\begin{figure}
\includegraphics[width=0.5\textwidth, angle=0]{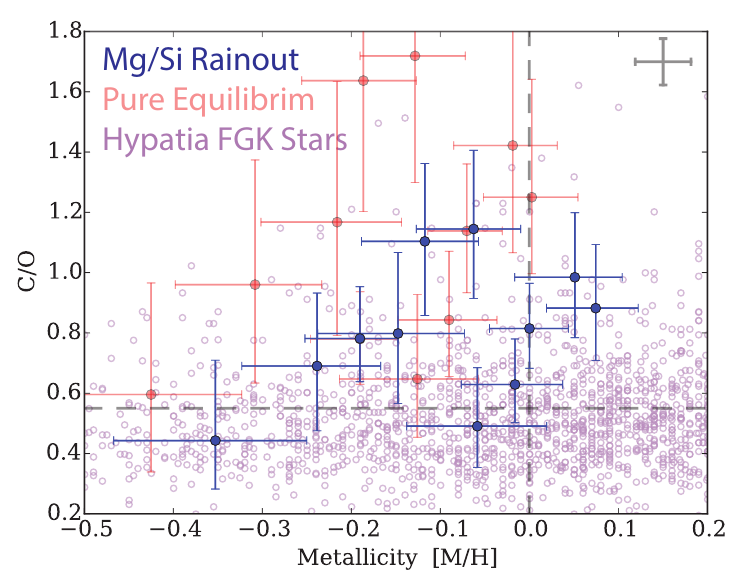}
     \caption{ \label{fig:figure6}  Bulk metallicity (sum of all elements heavier than hydrogen and helium) and carbon-to-oxygen ratios derived from the retrieved methane and water abundances.  The red points show those quantities derived from pure equilibrium. The dark blue points show those quantities derived assuming silicate rainout condensate chemistry by removing 3.28 oxygen atoms per silicon atom.  The data in purple was determined from the Hypatia Catalog (Hinkel et al. 2014), an amalgamate dataset of FGK stellar abundances, where all literature source measurements are solar normalized to Lodders et al. (2009).  The error bar in the top right represents the average spread in the abundance determinations for any given star (Hinkel et al. 2016). {   For our sample of late-T dwarfs, we find lower metallicities and higher C/O ratios than that of the average stellar population.} }     
\end{figure}

 Figure \ref{fig:figure6} summarizes the results of this analysis.  {   We derive sub-solar metallicity and super solar C/O for the majority of the late T dwarfs studied in this investigation}. Given that the retrieved water abundances for all objects is greater than or equal to the methane abundance, it is unsurprising that we in turn derive a super solar C/O.   We also find that the rainout assumption can have a fairly significant effect on the derived C/O and metallicity. Increasing the amount of oxygen lost because of rainout ``shifts" the population towards more solar-like abundances.  One could imagine that if an even more efficient mechanism for oxygen sequestration existed (e.g.,formation of more highly hydrated minerals, e.g., Montmorillonite, than predicted by equilibrium chemistry), {   the abundances in these objects would be further shifted towards} the solar values.  
 
In Figure \ref{fig:figure6} we compare the brown dwarf metallicities and carbon-to-oxygen ratios to those of the local FGK stellar population derived from the Hypatia catalogue (Hinkel et al. 2014).  Stellar abundance determinations are no trivial task and often different datasets and methodologies can lead to inconsistent results (e.g., Hinkel et al. 2016), resulting in relatively large systematic abundance uncertainties per any given star.  In general, we find that the objects in our sample tend to lie towards lower metallicities and higher carbon-to-oxygen ratios than the bulk of the stellar population. The reason for this is unknown, but we again hypothesize that it may have to due with how efficient oxygen is sequestered into condensates, preventing us from determining the true oxygen inventory. More detailed thermochemical condensate modeling is required to identify additional sinks of oxygen.  

{   It is also possible that an object could have been formed with solar metallicity and ended up with a non-solar Si/O ratio (e.g., Hinkel et al. 2014), which could alter the amount of oxygen sequestered in silicates}.  With a higher Si/O, more oxygen could be depleted through silicate rainout.  Both [Si/Fe] and [O/Fe] are known to decrease with decreasing metallicity ([Fe/H]) in the stellar population (Figure 5 \& 7 in Hinkel et al. 2014) because they are $\alpha$-elements (e.g., Gratton \& Ortolani 1986; Marcolini et al. 2009).  However, oxygen tends to decrease more than silicon with decreasing [Fe/H], suggesting that lower metallicity objects should have a {\it greater} proportion of O relative to Si than at higher metallicities.  A good reason as to why the intrinsic Si/O might be preferentially higher amongst brown dwarfs is currently unknown.  

\subsection{Clouds}\label{sec:clouds}
As demonstrated in the preceding sections (\S\ref{sec:element_abund} and \S\ref{sec:trends}), condensates are likely responsible for the depletion of oxygen and the alkali metal trend.  We might anticipate the resulting clouds to impact the {   emergent} spectrum.  As such, we have included a non-scattering gray cloud in our retrieval scheme (equation \ref{eq:cloud_eq}).  {   To investigate the importance of clouds we can reconstruct the cloud optical depth ($\tau_{cloud}$} \footnote{ {   $\tau_{cloud}=\int_{P_0}^{0}\kappa_{c}(P)\frac{dP}{g}=\int_{P_0}^{0}\kappa_{P_0}\left(\frac{P}{P_0}\right)^{\alpha}\frac{dP}{g}$. We choose cloud optical depth as our cloud metric as the other parameters in themselves are not particularly instructive at face value.}}, {  , Figure \ref{fig:figure3}) derived from the posteriors of the three retrieved cloud parameters (cloud base pressure ($P_{0}$), cloud base opacity ($\kappa_{P_0}$), cloud profile shape factor ($\alpha$))} in equation \ref{eq:cloud_eq}.   The cloud optical depth histograms for each object all show a dramatic decline in probability near optical depths of unity (vertical dotted lines). This suggests that for these objects, the data indicates that optically thick clouds are not present. This is consistent with our findings in Part I where we applied Bayesian hypothesis testing to show that the inclusion of clouds for Gl570D and HD3651B was not justified given the data nor did their inclusion impact the retrieved values for any of the other parameters.  We note that for some objects there is a small trickle of probability towards higher optical depths, though not significant relative to the total integrated probability.

{   The lack of large cloud optical depth agrees with the widely recognized expectation that late T-dwarfs should have relatively clear atmospheres over the pressure levels probed by the near-inrared (Lodders 1999; Lodders \& Fegley 2002; Visscher et al. 2006;2010; Morley et al. 2012)}. Based upon the intersection of the enstatite condensation curve with the TP-profiles (Figure \ref{fig:figure4}), we find that any {   enstatite} clouds should largely reside at the deepest pressure levels unseen by the near-infrared spectra, e.g., the gaseous opacities become optically thick at pressure levels above where the enstatite cloud should form and thus should have no impact on the emergent spectrum.  This is consistent with the findings of Morley et al. (2012), where forward models of objects covering a similar temperature range suggest that the {   enstatite} clouds largely reside below at pressure levels deeper than those probed by near-infrared observations, except for the objects approaching $\sim$900K.

Figure \ref{fig:figure4} and the alkali metal trend (described in detail in \S\ref{sec:trends}) suggests that Na$_2$S and KCl clouds should form within the region of atmosphere probed by the near-infrared spectra.  Morley et al. (2012) showed that the largely neglected cloud species Na$_2$S, KCl, MnS, and ZnS can form within cool dwarf atmospheres.  They showed for the latest T-dwarfs that, depending on gravity and the degree of droplet sedimentation (higher gravity and more vertically extended clouds result in higher cloud optical depths), the optical depths of Na$_2$S (more so than KCl) can range from negligible to a few over over wavelengths between 1 and 6$\mu$m (their Figure 4 and 8). Our photospheric optical depth upper limits  suggest that the cloud optical depths cannot be significantly greater than unity\footnote{However, we do not include multiple scattering, that is, we assume that the clouds are purely absorbing. It is therefore possible that there exist higher optical depth clouds, with non-zero single scatter albedo's ($\omega$), which would allow for higher optical depths by a factor of $\sim$(1-$\omega$)$^{1/2}$ for isotropic scattering }. This may potentially place constraints on the vertical extent of the {   Na$_2$S and KCl} clouds, largely favoring vertically compact clouds (thinner, higher degree of sedimentation within their modeling framework).  

{   Our derived cloud optical depth upper limits rule out optically thick clouds but cannot rule out optically thin clouds. From the aforementioned modeling studies, Na$_2$S and KCl condensates may form optically thin clouds as opposed to optically thick clouds. The existence of optically thin alkali clouds is consistent with our alkali abundance trend. A plausible physical picture is that the alkalis are condensing and rapidly coalescing into larger particles which sink quickly, resulting in optically thin clouds. In short, our results are internally self-consistent with regards to the alkali trend and lack of optically thick clouds}. Continued work on determining more precise cloud optical depth upper limits and bounded constraints on cloud optical depths will be invaluable in constraining cloud physics in substellar atmospheres. 

\subsection{Evolutionary Diagnostics: Radius, Gravity,  Effective Temperature, \& Age}\label{sec:Evo}
The radius, gravity, and effective temperature of a substellar object are diagnostic of its {   age, mass, and evolutionary history}. (Burrows et al. 1997; Chabrier et al. 2000; Burrows et al. 2001;   Baraffe et al. 2002; Saumon \& Marley 2008). {   The population of late T-dwarfs found in the field is expected to be dominated by older, higher gravity (log$g\sim$5), cool objects (Saumon \& Marley 2008).}   Gravities are not typically expected to exceed $\sim$5.3 even for the oldest, most massive objects (e.g., Saumon \& Marley 2008). 

{   In this work the effective temperature is not a directly obtainable parameter as it is not a free parameter like in grid model fits. Rather our effective temperature ($T_{eff}$) is derived by equating Boltzmann's law to the bolometric fluxes (1-20 microns) derived from 5000 model spectra drawn from the posterior.  The photometric radius ($R$) is derived from the retrieved $(R/D)^2$ scaling parameter given the parallax distances ($D$).}

\begin{figure}
\includegraphics[width=0.5\textwidth, angle=0]{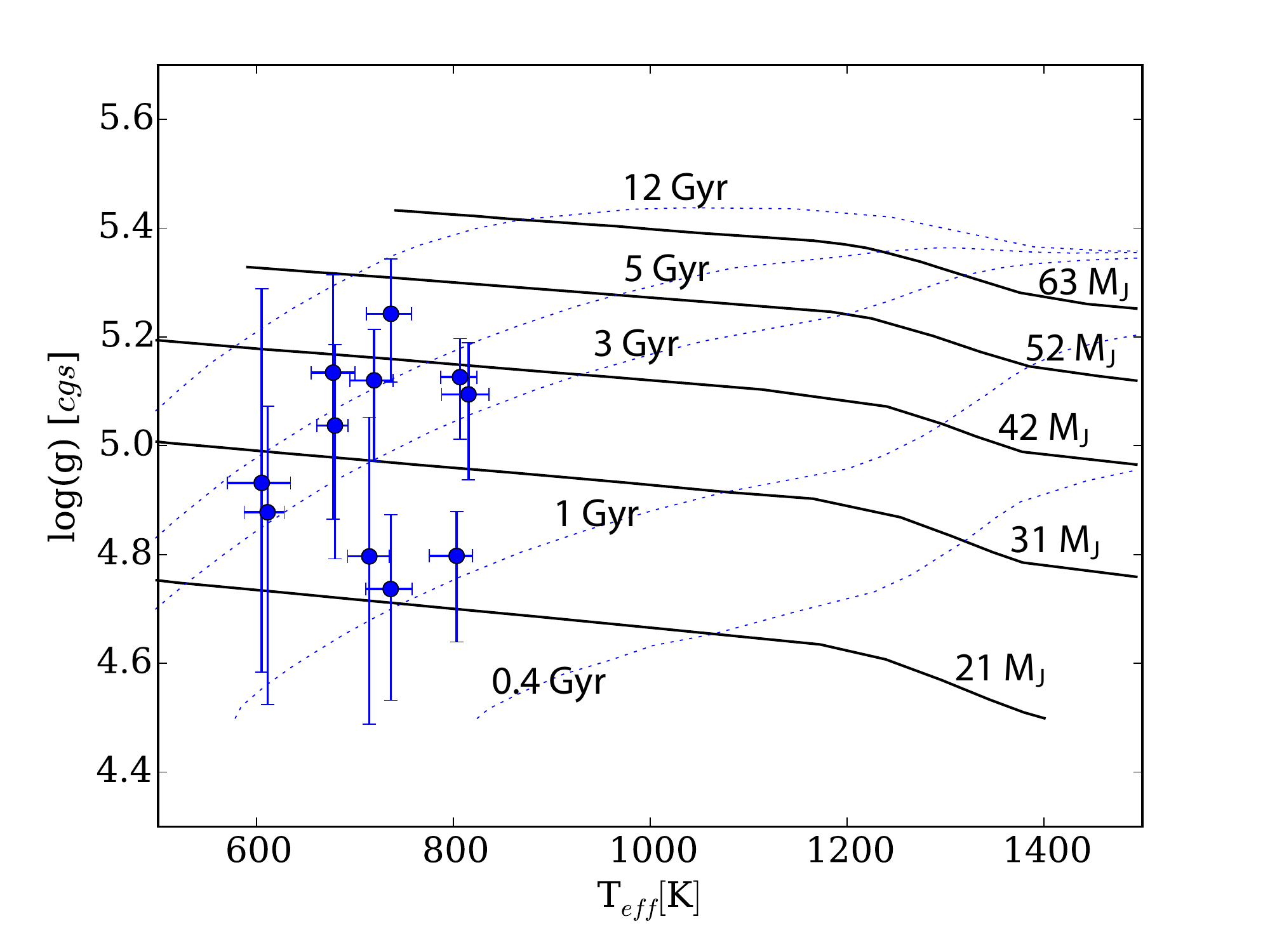}
     \caption{ \label{fig:figure9}  Comparison of retrieved gravities and effective temperatures with evolution models (Saumon \& Marley 2008, adapted from Part I).  The inferred ages (blue dashed curves) and masses (black curves) are consistent with expectations for field T-dwarfs.    }
  \end{figure}

Figure \ref{fig:figure9} summarizes the effective temperatures and gravities of our 11 objects compared with the evolution models of Saumon \& Marley (2008) (see also figure \ref{fig:figure3} derived with the Burrows 2001 analytic relations).  We find that the retrieved log-gravities of the objects fall between 4.75 and 5.3. Figure \ref{fig:figure9} shows that the retrieved gravity and derived effective temperatures are consistent with ages between 1 and 5 Gyr and masses between $\sim$20 and 50 Jupiter radii.  As an additional, more quantitative check, we use the approximate evolutionary relationships (as a function of gravity and effective temperature) from Burrows et al. (2001) (their equation 4)  to derive the possible ages for each object (last column in Figure \ref{fig:figure3}). Again, we find that the median ages for each object are typically greater than $\sim$1 Gyr, but some span a wide range of possible ages due to the larger uncertainties on gravity.  The gravity, effective temperatures, and ages of this sample are also in line with expectations from population synthesis evolutionary models (Figure  9 in Saumon \& Marley 2008) for late T-dwarfs and observations of mass-calibrated binary systems (Dupuy \& Liu 2017).

We find that the radii for most of the objects in our sample fall between $\sim$0.7 and 1.3 Jupiter radii (Figure \ref{fig:figure3}).  We can test for the self-consistency of the photometrically derived radii using the analytic evolutionary relationship of radius with effective temperature and gravity (equation 5 in Burrows et al. 2001)\footnote{As a check, we compared the radii derived from the Burrows et al. (2001) equations to the ``hybrid evolution grid" from Saumon \& Marley (2008) for relevant select T$_{eff}$-log$g$ pairs. We found that for the same T$_{eff}$-log$g$ pair, radii differences between the two models are no larger than 0.02R$_J$. This is small compared to the relative to the retrieved photometric radii errors of $\sim$0.08$R_J$.}. Propagating the uncertainties in gravities and effective temperatures for each object into the evolutionary relationship for radius we find that our photometrically derived radii are systematically higher\footnote{We note that the opposite is true for self-luminous directly imaged planets, as their photometrically derived radii tend to be systematically smaller than predicted from evolution models (e.g., Marley et al. 2012), possibly due to model composition assumptions.} than expected from the evolution models (Figure \ref{fig:figure3}).  

The largest discrepancies (in terms of differences in the medians of the radii distributions) come from 2MASS J1553, 2MASS J1217, and to a lesser degree 2MASS J0939.  As discussed in \S\ref{sec:Sample} 2MASS J1553 has been shown to be a binary,  whereas the evidence for 2MASS J1217 and 2MASS J0939 being binaries is more circumstantial.  In future work if we a priori know the object is a binary, or like in previous works (e.g., Burgasser et al. 2008) we identify an overly large radius, we may try fitting for the atmospheric properties of two separate objects and combining them into one spectrum (as done in Leggett et al. 2009).  At this time it is unclear why the evolution derived radii and the photometric radii are inconsistent.  Perhaps some of these other objects are indeed like 2MASS J1217 and 2MASS J0939 in they are unresolved binary systems (Aberasturi et al. 2014) .  

We note that the recent detection of a highly irradiated (T$_{eq}\sim$2400K) transiting brown dwarf, KELT-1b (Siverd et al. 2012, see also Bouchy et al. 2011), also shows a larger radius, relative to evolution models, of $1.116^{+0.038}_{-0.029}$R$_{J}$ given its mass of $\sim$27$M_{J}$ and age ($\sim$1.5 - 2 Gyr) .  However, these anomalously large radii may depend on high degree of stellar irradiation, driving some as-of-yet unknown inflation mechanism operating on many hot Jupiter's (e.g., Laughlin et al. 2011).  It is currently unclear how stellar irradiation driven inflation mechanisms would impact higher gravity objects (log$g$=4 - 5,  transiting planets have log$g$ $<$ 3.5 with most less than 3.).  However, Bouchy et al. (2011) argue that large stellar insolation is not likely to significantly inflate the radii of irradiated brown dwarf-mass objects.  These measured radii are typically $\sim$10\% larger than expectations from evolution models given their masses and a reasonable range of ages. The large radii for the non-binary systems in our sample are in good agreement with these transiting systems. However, some transiting brown dwarf mass objects, e.g., LHS 6343ABb which is not highly irradiated (Johnson et al. 2011; Montet et al. 2016), do not show evidence for inflated radii ($0.833^{+0.021}_{-0.021}$ R$_{J}$ ) and are in line with evolution model expectations.  Continuing to identify these benchmark systems over a range of irradiation levels will further test evolution and {   atmosphere} models and provide context for our radius determinations from retrieval analyses.

\section{Comparison with Previous Analysis}\label{sec:comp}
\begin{figure*}
\includegraphics[width=1\textwidth, angle=0]{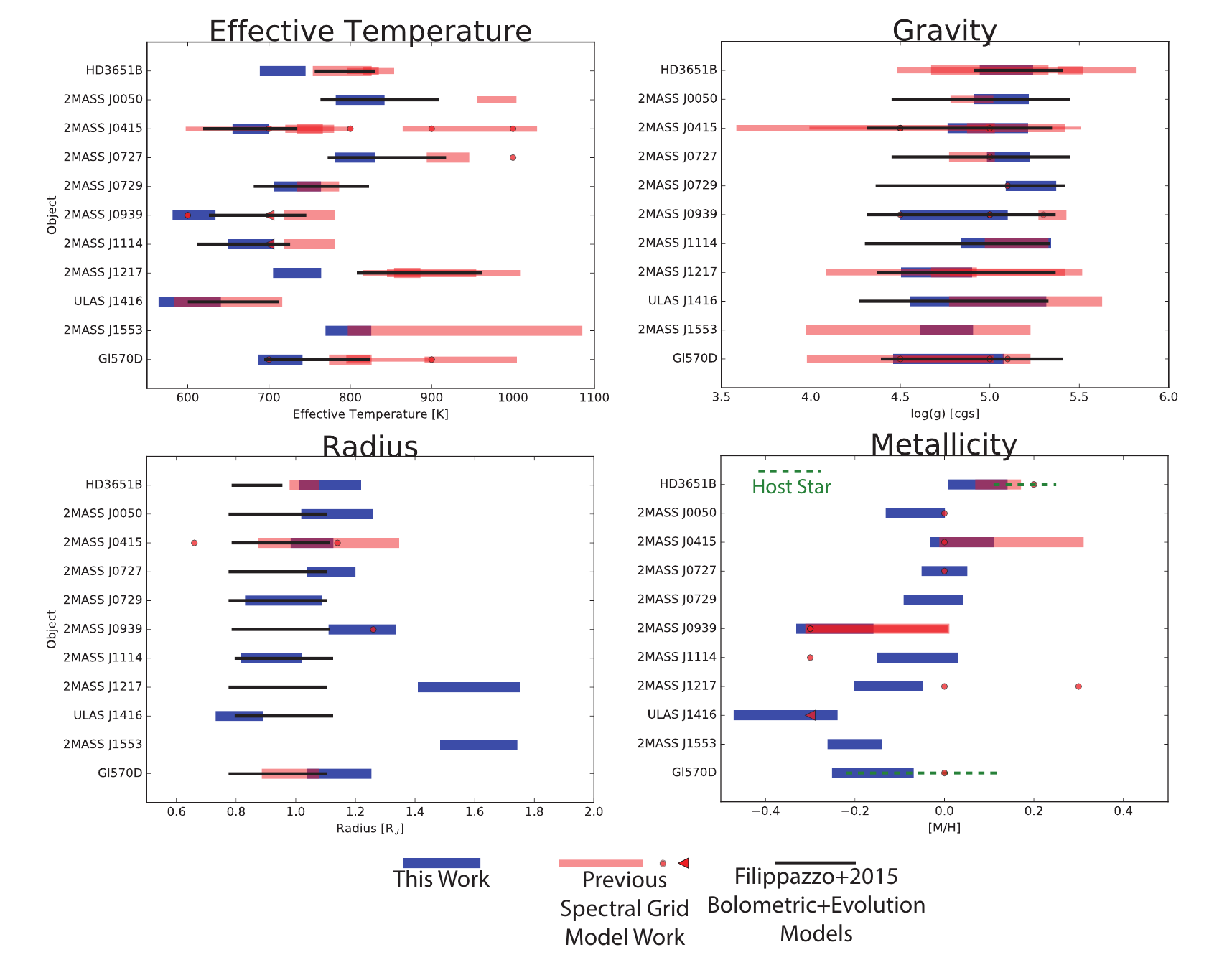}
     \caption{ \label{fig:figure10}  Summary of our retrieved quantities (from the data in Table \ref{tab:table3} and Table \ref{tab:table4} ) to those derived in the literature.  Our derived quantities (68\% confidence intervals) are shown as the blue bar for each object. We re-iterate that in our analysis, all but gravity are derived from the retrieval parameters as described in the text (effective temperature via integration of spectra drawn from the posterior, metallicity derived from the molecular abundances, and radius from the radius-to-distance dilution factor and parallax).  The spectroscopically derived literature values are shown as the red horizontal bars. Different thicknesses correspond to different studies.  Some literature values only have upper limits (red triangle) or point estimates (red circle).  In black we summarize the bolometric luminosity+evolution model derived quantities from Filippazzo et al. 2015.   Finally, in the lower right panel, we show the host star metallicities for the benchmark systems Gl570D and HD3651b presented in Line et al. 2015 (Part I) as the green dashed line. In general, we find that the Filippazzo et al. 2015 effective temperatures are cooler than the spectroscopic+grid model derived parameters, and our results lower still.  Gravities are all in near perfect agreement (albeit with large uncertainties).  Our radii sit somewhat higher than most, and our metallicities are typically in good agreement with the literature.     }
  \end{figure*}

Table \ref{tab:table3}, Table \ref{tab:table4}, and Figure \ref{fig:figure10} summarize our results compared with parameters from the literature derived via a variety of methods.  Table \ref{tab:table3} summarizes spectroscopically derived quantities from the literature based upon spectra-grid model comparisons, and Table \ref{tab:table4} compares our results to the bolometric luminosity plus evolution model results described in Filippazzo et al. 2015.   There are a wide range of literature values for effective temperatures, gravities, and radii for any individual object. In a broad sense, our results are largely in line with what has been found before.  However, we do find that in most cases our effective temperatures are systematically lower by up to $\sim$100K when compared to previous determinations. However, the differences in the estimates from previous works can also be on the order of one to several hundred Kelvin due to different wavelength regions explored and/or different model grids. Overall, we find that our derived effective temperatures are systematically the coolest (Figure \ref{fig:figure10}).  

Our gravity determinations agree very well with previous estimates, with most previous estimates for each object falling within our 68\% confidence interval (typically $\sim$ 0.2 dex).  The inferred metallicities are also typically in good agreement with previous determination, for select objects. 

 For objects previously reported to have low metallicities (2MASS J0939 and ULAS J1416), we too find significantly low metallicity.    We find a slightly enhanced, though not significantly, metallicity for 2MASS J0415, consistent with the 0.0-0.3 dex determination from Saumon et al. (2007).  We also find that the photometric radii are consistent with previous determinations where available (2MASS J0415 with Yamamura et al. (2010) reporting 1.14R$_{J}$ based on AKARI data is within 2-sigma of our value of 1.06$^{+0.05}_{-0.06}$R$_{J}$   ) and that binaries stand out as objects with overly large photometric radii (e.g.,  1.26R$_{J}$ for 2MASS J0939 from Burgasser et al. 2008 ,  compared with our determination of 1.22$^{+0.1}_{-0.09}$R$_{J}$ ). 

{   Filippazzo et al. 2015 (Table \ref{tab:table4}) reconstructed the bolometric fluxes from available spectroscopic and broadband data.  From the bolometric fluxes and the parallactic distances, they were able to determine the bolometric luminosities.  From the luminosities combined with the evolutionary model grids of Saumon \& Marley 2008, Chabrier et al. 2000, and Baraffe et al. 2003, they were able to infer gravites, effective temperatures, and radii assuming typical field ages (0.5 - 10 Gyr).  We find reasonable agreement with the Filippazzo et al. (2015) results. Our bolometric luminosities are typically higher than theirs, but are almost always within 2$\sigma$.  The larger luminosities can be explained by the larger photometric radii we retrieve. }

\begin{table*}
\centering
\caption{\label{tab:table4} Summary of Evolutionary model derived parameters from Fillappezzo et al. 2015 (F10) compared with our retrieval results. }
\begin{tabular}{lccccl}
\hline
\hline
\cline{1-2}
 Object             &                     \Teff    &      \logg    &     $R$   &   log(L$_{\star}$/L$_{\sun}$) & Ref  \\
               &                              (K)    &     (cgs)     &    ($R_{J}$)   &  &    \\
\hline
        
HD3651B        			         &   793$^{+35}_{-35}$   & 5.16$^{+0.24}_{-0.24}$ &  0.86$^{+0.07}_{-0.07}$   & -5.56$^{+0.01}_{-0.01}$   & F10  \\     
						&   719$^{+19}_{-25}$   & 5.12$^{+0.1}_{-0.2}$ &  1.10$^{+0.1}_{-0.07}$   &-5.51$^{+0.05}_{-0.05}$ & This Work  \\  
\\						
2MASS J00501994$-$3322402         &   836$^{+71}_{-71}$   & 4.95$^{+0.49}_{-0.49}$ &  0.94$^{+0.16}_{-0.16}$   & -5.39$^{+0.02}_{-0.02}$   & F10  \\ 
                                			         &  815$^{+20}_{-27}$   & 5.09$^{+0.1}_{-0.2}$  &  1.12$^{+0.12}_{-0.09}$   & -5.27$^{+0.06}_{-0.06}$ & This Work  \\          
\\			         
	2MASSI J0415195$-$093506          &   677$^{+56}_{-56}$   & 4.83$^{+0.51}_{-0.51}$ &  0.95$^{+0.16}_{-0.16}$   & -5.74$^{+0.01}_{-0.01}$   & F10  \\ 
			         			      &  680$^{+13}_{-18}$   & 5.04$^{+0.2}_{-0.2}$  &  1.06$^{+0.05}_{-0.06}$ & -5.70$^{+0.04}_{-0.04}$   & This Work  \\        
\\						      
 2MASSI J0727182+171001            &   845$^{+71}_{-71}$   & 4.95$^{+0.49}_{-0.49}$ &  0.94$^{+0.16}_{-0.16}$   & -5.37$^{+0.01}_{-0.01}$   & F10  \\           
			          &  807$^{+17}_{-19}$   & 5.13$^{+0.1}_{-0.1}$    &  1.12$^{+0.07}_{-0.06}$ & -5.30$^{+0.03}_{-0.03}$   & This Work  \\                                                                  
 \\                                                                                                    
2MASS J07290002$-$3954043         &   752$^{+69}_{-69}$   & 4.89$^{+0.52}_{-0.52}$ &  0.94$^{+0.16}_{-0.16}$   & -5.57$^{+0.06}_{-0.06}$   & F10  \\
			        &  737$^{+21}_{-25}$   & 5.29$^{+0.1}_{-0.1}$   &  0.95$^{+0.12}_{-0.10}$ & -5.60$^{+0.08}_{-0.08}$  & This Work  \\                                                                  
                                                                                                     
  \\                                                   
 2MASS J09393548$-$2448279       &   686$^{+58}_{-58}$   & 4.84$^{+0.52}_{-0.52}$ &  0.95$^{+0.16}_{-0.16}$   & -5.72$^{+0.02}_{-0.02}$   & F10  \\                                                     
			          &  611$^{+17}_{-24}$   & 4.88$^{+0.2}_{-0.4}$    &  1.22$^{+0.1}_{-0.09}$&  -5.71$^{+0.07}_{-0.07}$  & This Work  \\

 \\                                                                                                    
2MASS J11145133$-$2618235          &   669$^{+55}_{-55}$   & 4.82$^{+0.51}_{-0.51}$ &  0.96$^{+0.16}_{-0.16}$   & -5.76$^{+0.01}_{-0.01}$   & F10  \\    
			         &  678$^{+22}_{-22}$   & 5.13$^{+0.2}_{-0.3}$ &  0.91$^{+0.09}_{-0.08}$ & -5.77$^{+0.05}_{-0.05}$  & This Work  \\                                                                  
 \\                                                                                                    
2MASSI J1217110$-$031113         &   885$^{+75}_{-75}$   & 4.87$^{+0.49}_{-0.49}$ &  0.94$^{+0.16}_{-0.16}$   & -5.29$^{+0.02}_{-0.02}$   & F10  \\   
			         &  726$^{+22}_{-25}$   & 4.74$^{+0.1}_{-0.2}$  &  1.57$^{+0.11}_{-0.22}$ &  -5.16$^{+0.05}_{-0.05}$  & This Work  \\                                                                  
 \\                                                                                                    
ULAS J141623.94+134836.3         &   656$^{+54}_{-54}$   & 4.80$^{+0.52}_{-0.52}$ &  0.96$^{+0.16}_{-0.16}$   & -5.79$^{+0.01}_{-0.01}$   & F10  \\               
			         &  605$^{+29}_{-35}$   & 4.93$^{+0.4}_{-0.4}$   &  0.8$^{+0.07}_{-0.06}$ & -6.08$^{+0.08}_{-0.08}$  & This Work  \\         
\\			                                                             		         
Gliese 570D        &  759$^{+63}_{-63}$   & 4.90$^{+0.5}_{-0.5}$  &  0.94$^{+0.16}_{-0.16}$   & -5.55$^{+0.01}_{-0.01}$ & F10   \\
        &  715$^{+20}_{-22}$   & 4.80$^{+0.3}_{-0.3}$  &  1.14$^{+0.1}_{-0.09}$   & -5.49$^{+0.04}_{-0.05}$  & This Work  \\

\hline
\end{tabular}
\end{table*}

\section{Summary \& Conclusions}\label{sec:Conclusions}
We have applied the benchmark-validated retrieval methodologies of Line et al. 2015 (Part I) to low-resolution near-infrared spectra of 11 late-T dwarfs. A uniform analysis of objects observed with the same instrument, and interpreted using the same modeling tools provides a solid foundation for atmospheric property comparison and identification of diagnostic trends.   


From this analysis, we have arrived at the following conclusions: \\
\begin{enumerate}
\item Our retrieval modeling approach can fit the data very well. The residuals are quite small and appear to be random (Figure \ref{fig:figure1}). We can, in general, fit the data better than ultra-cool dwarf gird models (Liu et al. 2011; Morley et al. 2012), and our large number of model fitting parameters is statistically justified (Figure \ref{fig:figure0}, \S\ref{sec:intro}).

\item The retrieved temperature-pressure profiles are consistent with expectations from radiative-convective equilibrium temperature profiles over the pressure levels most densely probed by the near-infrared spectra (Figure \ref{fig:figure4}). Deviations from grid model expectations which would be potentially diagnostic of additional physics, that occur at the lowest or highest pressure levels should not be too heavily interpreted (\S\ref{sec:TP}).

\item The mixing ratios of only water, methane, ammonia, and the alkalis have strong, bounded constraints (mole fractions typically constrained at 68\% confidence to within $\le$0.2 dex, or a factor of  1.6).  Ammonia is present and well constrained in all objects, consistent with strong detections presented via three independent methods in Part I.  The retrieved molecular abundances for these species are largely consistent with thermochemical equilibrium predictions for given temperature-pressure profile, metallicity, and carbon-to-oxygen ratio. This suggests chemical plausibility (Figure \ref{fig:figure5}, \ref{sec:chemistry}).   We only obtain upper limits on the mixing ratios of the other included species (CO, CO$_2$, and H$_2$S) (Figure \ref{fig:figure3}, \S\ref{sec:comp}). 

\item There is no detectable trend in the mixing ratios of methane, water, or ammonia with effective temperature. There is, however, a very strong trend in the alkali metal mixing ratios with effective temperature (Figure \ref{fig:figure7}).   This alkali trend is consistent with expectations from condensate rainout chemistry, which results in a depletion of the column integrated mixing ratios of elemental sodium and potassium at cooler effective temperatures as the temperature-condensate intersection pressures for Na$_2$S and KCl move towards the deeper atmosphere (Figure \ref{fig:figure8}, \S\ref{sec:trends}).

\item Our elemental abundance analysis of the retrieved methane and water mixing ratios suggests sub-solar metallicities and somewhat super-solar carbon-to-oxygen ratios, which are somewhat lower and higher, respectively, when compared to the field FGK stellar abundances (\S\ref{sec:element_abund}).  One caveat is that the late T-dwarf metallicities and carbon-to-oxygen ratios are sensitive to the degree of oxygen depletion in condensates that form in the deep atmosphere (Figure \ref{fig:figure6}).  More efficient oxygen depletion due to additional rainout mechanisms could shift the derived metallicity and carbon-to-oxygen ratios towards solar.

\item We only retrieve cloud optical depth upper limits (Figure \ref{fig:figure3}) . These upper limits are typically near unit optical depth suggesting that clouds do not substantially influence the spectra of these 11 late T-dwarfs (\S\ref{sec:clouds}). This is consistent with our findings from Part I where the inclusion of cloud parameters was not justified given the data, nor did their inclusion, or lack thereof, influence in any way the constraints for the other parameters.

\item The retrieved gravities and derived effective temperatures are largely consistent with evolutionary models (Figure \ref{fig:figure9}) and previous works, though our retrieved effective temperatures are systematically cooler than previous grid-model derivations.   We also find that the photometrically derived radii of several of the objects are larger than those expected from the evolution models derived from the retrieved effective temperatures and gravities. Some of these large radii are indicative of binary systems, consistent with the literature. The other objects' larger radii are difficult to explain, but are in line with what has been found for some transiting brown dwarfs (\S\ref{sec:Evo}).

\end{enumerate}

We emphasize, that all of these constraints have been made with relatively low-resolution spectra (SpeX Prism, at R=80-120). High-resolution is not necessarily a requirement for obtaining useful abundance information, {   unlike in classic stellar abundance analysis}. Broad spectral coverage spanning multiple molecular bands goes a long way towards constraining molecular abundances, even with a large number of highly degenerate parameters.

\section{Discussion \& Future Work}
While we have learned a lot from just this small sample, there is still very much unknown and much work to be done.  Expanding our sample to more objects over a wider range of temperatures will allow us to better delineate the preliminary trends presented here. We would like to first expand our sample to more low resolution spectra of late T-dwarfs and Y-dwarfs. We have demonstrated that our methodologies work within this cool temperature, largely cloud-free, regime and for low-resolution data. We should understand these simpler objects before we can move onto more complex, warmer, cloudier objects (e.g., Burningham et al. accepted).  

As we extend our analysis to cooler temperatures, we would like to definitively identify the transition effective temperature at which we can no longer determine bounded constraints on the alkali abundances. Color-spectral type analysis suggest that this transition occurs around T8, near the Y-T transition (e.g., Liu et al. 2012; Schneider et al. 2015).  Measuring the transition abundance and effective temperature would place further constraints on the rainout mechanism and can be directly compared to equilibrium chemistry predictions.  

As we extend our analysis towards slightly higher temperatures, we would like to identify when the strong ammonia detection disappears (i.e., becomes an upper limit), or if we can retrieve a decline in its abundance with increasing temperature (e.g., Figure 5, Burrows \& Sharp 1998). The rate of decline\footnote{Ammonia is thermochemically expected to disappear around T2 and earlier. Vertical mixing could dredge up ammonia from more ammonia rich deeper atmospheric layers resulting in more abundant ammonia than anticipated at earlier spectral types. Determining the delay could help identify the average strength of vertical mixing. }, and at what effective temperature this transition occurs relative to equilibrium chemistry predictions, could possibly place constraints on the role of disequilibrium nitrogen chemistry in early T-dwarf atmospheres (e.g., the dispersion in vertical mixing strengths).  

{   As shown with our retrieval results, it is difficult to obtain reliable TP-profile information deep in the atmosphere due to the increasing significance of collision induced opacities.  Higher resolution spectra near the water band-heads (between the YJH bands) or longer wavelengths with higher opacity could allow us to probe {\it lower} pressures (higher altitudes) than probed here, permitting more precise upper atmosphere temperature constraints.  Hotter objects, mid T-s and earlier types, will present a challenge as the near infrared spectra probe a more narrow range of pressure levels due to the presence of clouds, preventing us from accessing the deeper atmospheric levels. Any temperature profile information below optically thick clouds will have to come from extrapolation via parameterizations or assumed from physically motivated models. } 

{   Continued validation of modeling approaches using benchmark systems over a range of mass ratios, like the stellar-brown dwarf systems Gl570 and HD3651 from Part 1, or brown-dwarf brown-dwarf systems like WISE-1217+1626AB (Liu et al. 2012; Leggett et al. 2014) or WISE-0146+4234AB (Dupuy et al. 2015),  is also necessary to identify improvements and/or pitfalls in our current modeling, and to understand the partitioning of elements as a function of mass-ratio}.

 In that regards, there are several possible avenues of modeling improvements. Firstly, more testing and identification of robust temperature-pressure profile parameterizations are needed.  Our current ``p-spline smoothing prior" parameterization introduced in Part I works well for late T-dwarfs, and is in agreement with expectations from self-consistent cloud-free grid models. However, cloudy objects with limited vertical dynamic range will require less flexible parameterizations with fewer free parameters (e.g., Burningham et al. accepted). Understanding the influence of these temperature profile assumptions on the abundances, gravities, cloud optical depths, etc., will become more important as we move towards hotter objects or objects with less spectral coverage and signal-to-noise, like self-luminous directly imaged planets.   The role that various cloud parameterizations play in influencing the other retrieved properties will also need to be explored (e.g., gray non-scattering, vs. wavelength dependent scattering, cloud vertical profiles/particle sizes/compositions/patchiness, etc.), as has been done within grid-modeling frameworks (e.g., Burrows et al. 2006; Cushing et al. 2008; Morley et al. 2012).  Finally, more sophisticated data-model comparison techniques should be explored (e.g., Czekala et al. 2015) especially when retrieving on higher resolution spectra and stitching together spectra from different instruments, orders, and epochs.  Failure to account for these subtleties may result in biased answers.  

Despite these challenges, we are optimistic that continued improvement and application of retrieval methodologies to large samples of objects spanning a wide array of temperatures, gravities, and ages, will undoubtably provide new insights and surprises in physics and chemistry of brown dwarf atmospheres.
  
\section{Acknowledgements}
We thank Roxana Lupu and Richard Freedman for continued development and support of their extensive opacity database of which makes much of this work possible.  We also thank Michael Cushing, Jackie Faherty, Trent Dupuy, Brenden Bowler, Didier Saumon, Ryan Garland, Ty Robinson \& Adam Schneider for useful discussions. We also thank Dan Foreman-Mackey for his publicly available EMCEE code and the useful plotting routine corner.py.  We are also generally thankful to the python/matplotlib/numpy/scipy developers.  M.R.L. acknowledges support provided by NASA through Hubble Fellowship grant 51362 awarded by the Space Telescope Science Institute, which is operated by the Association of Universities for Research in Astronomy, Inc., for NASA, under the contract NAS 5-26555 and NSF grant Grant No. 1615220. MCL acknowledges support from NSF AST-1518339. BB acknowledges financial support from the European Commission in the form of a Marie Curie International Outgoing Fellowship (PIOF-GA-2013- 629435). NRH acknowledges the support of the Vanderbilt Office of the Provost through the Vanderbilt Initiative in Data-intensive Astrophysics (VIDA) fellowship.

\end{document}